\documentclass[fleqn,usenatbib]{mnras}
\usepackage{newtxtext,newtxmath}

\usepackage[T1]{fontenc}
\usepackage{ae,aecompl}

\usepackage{graphicx}   % Including figure files
\usepackage{amsmath}    % Advanced maths commands
\usepackage{threeparttable} %table with notes
\usepackage{threeparttablex} %longtable with notes

\title[Observations of SN~2002gh]{A puzzle solved after two decades: SN~2002gh among the brightest of superluminous supernovae}
\author[Cartier et al.]{R\'egis Cartier,$^{1}$\thanks{regis.cartier@noirlab.edu}
Mario Hamuy,$^{2,3}$ Carlos Contreras,$^{4}$ Joseph\,P. Anderson,$^{5}$ 
\newauthor Mark\,M. Phillips,$^{4}$ Nidia Morrell,$^{4}$  Maximilian\,D. Stritzinger,$^{6}$ Emilio D. Hueichapan,$^{1}$ 
\newauthor Alejandro Clocchiatti,$^{7,8}$ Miguel Roth,$^{4}$ Joanna Thomas-Osip,$^{9}$ and Luis E. Gonz\'alez$^{10}$ \\ 
$^{1}$Cerro Tololo Inter-American Observatory, NSF's National Optical-Infrared Astronomy Research Laboratory, Casilla 603, La Serena, Chile \\
$^{2}$Fundaci\'on Chilena de Astronom\'ia, Santiago, Chile \\
$^{3}$Hagler Institute for Advanced Studies, Texas A\&M University, Texas, USA\\
$^{4}$Las Campanas Observatory, Carnegie Observatories, Casilla 601, La Serena, Chile\\
$^{5}$European Southern Observatory, Alonso de C\'ordova 3107, Casilla 19, Santiago, Chile \\
$^{6}$Department of Physics and Astronomy, Aarhus University, Ny Munkegade 120, 8000 Aarhus C, Denmark\\
$^{7}$Instituto de Astrof\'isica, Facultad de F\'isica, Pontificia Universidad Cat\'olica de Chile, Av. Vicu\~na Mackenna 4860, Santiago, Chile\\
$^{8}$Millennium Institute of Astrophysics, Nuncio Monse\~nor S\'otero Sanz 100, Providencia, Santiago, Chile\\
$^{9}$Gemini Observatory, NSF's National Optical-Infrared Astronomy Research Laboratory, Casilla 603, La Serena, Chile \\
$^{10}$Departamento de Astronom\'ia, Universidad de Chile, Casilla 36-D, Santiago, Chile\\
}

\begin{document}

\newcommand\fsec{\mbox{$.\!^{s}$}}%
\newcommand\arcdeg{\mbox{$^\circ$}}%

\newcommand\plottwo[2]{{% 
\typeout{Plottwo included the files #1 #2}
 \centering
 \leavevmode
 \columnwidth=.45\columnwidth
 \includegraphics[width={\eps@scaling\columnwidth}]{#1}%
 \hfil
 \includegraphics[width={\eps@scaling\columnwidth}]{#2}%
}}%

\newcommand{\vsi}{$v_{Si}$}
\newcommand{\vsidot}{$\dot{v}_{Si}$}
\newcommand{\kms}{km~s$^{-1}$}

\pagerange{\pageref{firstpage}--\pageref{lastpage}} \pubyear{2022}

\maketitle

\label{firstpage}

\begin{abstract}
We present optical photometry and spectroscopy of the superluminous SN\,2002gh
from maximum light to $+204$ days, obtained as part of the Carnegie Type\,II Supernova (CATS) project.
SN\,2002gh is among the most luminous discovered supernovae ever, yet it remained unnoticed
for nearly two decades. Using Dark Energy Camera archival images we identify the potential SN
host galaxy as a faint dwarf galaxy, presumably having low metallicity, and in an apparent merging
process with other nearby dwarf galaxies. We show that SN\,2002gh is among the brightest hydrogen-poor SLSNe with $M_{V} = -22.40 \pm 0.02$,
with an estimated peak bolometric luminosity of $2.6 \pm 0.1 \times 10^{44}$ erg s$^{-1}$.
We discount the decay of radioactive nickel as the main SN power mechanism, and assuming that the SN
is powered by the spin down of a magnetar we obtain two alternative solutions.
The first case, is characterized by significant magnetar power leakage,
and $M_{\mathrm{ej}}$ between 0.6 and 3.2 $M_{\sun}$, $P_{\mathrm{spin}} = 3.2$\,ms,
and $B = 5 \times 10^{13}$\,G. The second case does not require power leakage, resulting
in a huge ejecta mass of about 30 $M_{\sun}$, a fast spin period of $P_{\mathrm{spin}} \sim 1$\,ms,
and $B\sim 1.6 \times 10^{14}$\,G. We estimate a zero-age main-sequence mass 
between 14 and 25 $M_{\sun}$ for the first case and of about 135 $M_{\sun}$ for the second case.
The latter case would place the SN progenitor among the most massive stars observed
to explode as a SN.
\end{abstract}


\section{Introduction}

The first half of the 21$^{st}$ century will be remembered as a revolutionary epoch for time-domain astronomy.
Thanks to large area and untargeted transient surveys, astronomers are exploring the time-domain Universe
to an unprecedented level, discovering transient events that span a wide range of timescales and luminosities.
One of the most spectacular such transients are the rare class of supernovae (SNe) initially characterized by their
extremely bright peak luminosities ($\simeq -21$ mag at optical wavelengths) powered by a non-standard source
of energy, that have challenged our understanding of massive star evolution and explosion.  

These rare and bright SNe are commonly referred to as superluminous SNe (SLSNe). Like their lower luminosity
cousins they are divided into subtypes based on the presence of hydrogen in their spectra as hydrogen-rich
(Type II) and hydrogren-poor (Type I) \citep[see][]{moriya18, galyam19}.
Some SLSNe-II show relatively narrow hydrogen emission lines on top of a broad component, characteristic of
SNe Type IIn \citep[see][]{schlegel90}, and evidence of interaction between the
energetic SN ejecta and a massive hydrogen-rich circum-stellar medium (CSM), such as SN\,2006gy \citep{smith07, ofek07}.
These SLSNe-II are usually considered the bright end of Type IIn SNe, although it is unclear whether an additional power source
other than the strong ejecta-CSM interaction and radioactive decay
significantly contributes to their extreme luminosities \citep[see e.g.,][]{woosley07, jerkstrand20}. The idea of an additional
power source is reinforced by some SLSNe-II that do not show clear signatures of strong ejecta-CSM interaction \citep{gezari09, miller09, inserra18, dessart18}.
A potential power source is the energy injection from the spin down of a fast rotating neutron star with an extremely
powerful magnetic field, a magnetar, as explained below.  
Other notable events of the SLSN class, are a handful of objects initially classified as hydrogen-poor SLSNe, but
several days after maximum light reveal hydrogen features and signatures of ejecta-CSM interaction
\citep{yan15, yan17}, suggesting that some hydrogen-poor SLSNe
lose their hydrogen envelopes shortly before their final explosion.

Objects such as SN\,1999as \citep{1999IAUC.7128....1K}, SN\,2005ap \citep{quimby07}, SCP 06F6 \citep{barbary09},
and SN\,2007bi \citep{galyam09} were the first SLSN-I reported, several years before they were distinguished as a
separate SN class. Their bright peak luminosities together with the identification of a common set of spectral features
made it possible to distinguish them as members of a completely new SN class \citep{quimby11}.
These SLSNe are characterized by their extreme peak luminosities, very blue and
nearly featureless optical spectral emission, the lack of hydrogen and helium lines and the presence
of \ion{O}{ii} and \ion{C}{ii} absorption lines before and close to maximum light \citep{quimby11}.
A few weeks after maximum light, when the ejecta are cool enough, SLSN-I spectra become similar to SNe Ic
\citep[e.g.,][]{pastorello10} or to broad line SNe Ic \citep[SNe Ic-BL;][]{liu17} at an early phase.
Their nebular spectra also show some similarity with the nebular spectra of
SNe Ic-BL \citep{milisavljevic13,jerkstrand17,nicholl16b,nicholl19}. 

One of the first models proposed to account for the large luminosity displayed by SLSNe was the
radioactive decay of several solar masses of $^{56}$Ni \citep{quimby07, galyam09}, synthesized 
in a Pair Instability SN (PISN) \citep{barkat67, ravaky67}. 
PISNe are the theoretical explosions of low metallicity
and very massive main-sequence stars ($\sim 140-260$\,M$_{\sun}$), ejecting several tens of solar masses of $^{56}$Ni
\citep{heger02}. These objects are expected to be common in the early Universe. The large amount of radioactive material ejected by PISNe can
potentially power the extreme luminosities observed in SLSNe, however, some hydrogen-poor SLSNe fade too fast after peak to be consistent
with the $^{56}$Ni $\rightarrow$ $^{56}$Co $\rightarrow$ $^{56}$Fe decay chain, arguing against this power mechanism. Also the large opacity of iron-peak
elements is expected to shift the emission towards near-infrared wavelengths (NIR), which is in conflict with 
the very blue continuum displayed by hydrogen-poor SLSNe \citep{dessart12}. Additional evidence against hydrogen-poor
SLSNe being the result of PISNe, comes from the comparison between the observed spectra of SLSNe at nebular phases, which are dominated
by emission lines of intermediate mass elements \citep{milisavljevic13,jerkstrand17,nicholl16b,nicholl19,mazzali19}, in contrast with 
the iron-peak element dominated nebular emission predicted for PISNe \citep[see e.g.,][]{dessart12,jerkstrand16,mazzali19}.

An alternative model to explain the light curve evolution of hydrogen-poor SLSNe 
invokes the spin-down of a highly magnetic newborn neutron star, a ``magnetar'', that energises
the SN ejecta \citep{maeda07,woosley10, kasen10}. The magnetar model was introduced by \citet{maeda07}
to explain the peculiar double peaked and fast declining light curve of the Type Ib SN\,2005bf. After the recognition
of SLSNe as a new class, and using a small sample of well observed hydrogen-poor SLSNe,
\citet{inserra13} fitted an analytic magnetar model to their sample bolometric light curves and first showed that the power injection
from the spin down of a magnetar can successfully reproduce the complete light curve evolution, including a light curve flattening
at late times ($\sim 200$ days), that cannot be explained by the $^{56}$Co radioactive decay. More recently, \citet{nicholl17c}
presented a more sophisticaded version of the magnetar model built-in the Modular Open Source Fitter for Transients
\citep[{\sc MOSFiT};][]{guillochon18}. They applied their magnetar model to a large sample of hydrogen-poor SLSNe collected from
the literature and showed that such a magnetar model can explain the light curve evolution of many objects of this class.
Another alternative mechanism to power SLSN-I luminosity, is the SN ejecta-CSM interaction with a hydrogen and helium
free CSM. For complete recent reviews on the SLSNe properties and their power sources see \citet{galyam19} and
\citet{moriya18}.

Recently, the light curves of the hydrogen-poor SLSNe discovered in the course of the
Palomar Transient Factory \citep[PTF;][]{decia18}, the Pan-STARRS1 Medium Deep Survey
\citep[PS1;][]{lunnan18} and the Dark Energy Survey \citep[DES;][]{angus19}
were presented and analysed. These surveys show that the low-luminosity tail of the SLSN
population extends and overlaps in luminosity with the normal stripped envelope core
collapse SN (CCSN) population. Although there is a continuous distribution in brightness from SNe Ic, Ic-BL to SLSNe-I,
and rare examples of non-SLSNe powered by a magnetar exist \citep[e.g.,][]{maeda07, grayling21},
the dominant power mechanism seems to be different in the different SN populations.
The well-sampled multi-band light curves provided by these surveys, confirmed the large
diversity in the morphology of SLSN light curves, many of them showing bumps in their light curve
evolution \citep[see e.g.,][]{nicholl15,smith16,yan17}.

The host galaxies of hydrogen-poor SLSNe are characterized by being low mass dwarf galaxies
\citep[$M_{\mathrm{stellar}} < 2 \times 10^{9} \mathrm{M}_{\sun}$; e.g.,][]{neill11, lunnan14, leloudas15, angus16, perley16, schulze18},
usually of irregular morphology with roughly half of the galaxies exhibiting a morphology that is either asymmetric,
off-centre or consisting of multiple peaks \citep{lunnan15}, the latter being a possible signature of merging systems. 
SLSN host galaxies are clearly different from the galaxies hosting normal core-collapse SNe (CCSNe), which are more massive and in
general exhibit spiral structure. SLSN-I host galaxies are also characterized by low metallicities
\citep[$Z \leq 0.5 Z_{\sun}$; e.g.,][]{lunnan14, leloudas15, perley16, schulze18},
and by high specific Star Formation Rates \citep[sSFR $\sim 10^{-9}$ yr$^{-1}$; e.g.,][]{neill11,lunnan14,perley16}.
The host galaxies of SLSNe-II are more diverse, showing a range of properties varing from low mass and metal
poor dwarf galaxies, similar to the hosts of hydrogen-poor objects, to bright spiral galaxies similar to the
hosts of normal CCSNe \citep{leloudas15, perley16, schulze18}.

Here we present and analyse observations of SN\,2002gh obtained by the ``Carnegie
Type II Supernova Survey'' (CATS, hereafter). CATS was a SN follow-up
program similar to its successor the Carnegie Supernova Program \citep[CSP;][]{hamuy06},
carried out at Las Campanas Observatory during 2002–2003 with the main purpose
to study nearby ($z < 0.05$) SNe II. In the course of the CATS survey we included 
a handful of SNe of other classes. More details about the CATS program and the instruments
used can be found in \citet{hamuy06, hamuy09} and \citet{cartier14}. Remarkably,
CATS results include observations of SN\,2002ic, the first SN Ia showing interaction with a dense circumstellar
medium \citep{hamuy03}, and SN\,2003bg the first Type IIb hypernova \citep{hamuy09}.
Here we add to these notable achievements, the observations of SN\,2002gh, the second SLSN discovered after SN\,1999as
\citep{1999IAUC.7128....1K}, for which its superluminous nature remained unnoticed for nearly two
decades until now. In Section \ref{sec:observations} we describe the data reduction and present
the observations. In Section \ref{sec:spec_analysis}, we identify the defining SLSN-I spectral features
in the spectra of SN\,2002gh by comparing with other well studied objects, and measure line velocities. 
In Section \ref{sec:host_galaxy} we use Dark Energy Camera \citep{flaugher15} archival images to
identify and characterize the potential SN host galaxy. In Section \ref{sec:light_curve_chara} we characterize the light
curves of SN\,2002gh and in Section \ref{bolometric:sec} we use blackbody fits to estimate the bolometric peak
luminosity and the total radiated energy. We compare the bolometric light curve of SN\,2002gh 
with SLSNe from the PS1 and DES surveys, showing that SN\,2002gh is among the brightest
hydrogen-poor SLSNe. In Section \ref{power_source_sec} we explore the magnetar model and the
radioactive decay of $^{56}$Ni as potential power sources to explain the extreme luminosity
of SN\,2002gh, and discuss potential signatures of ejecta-CSM interaction. 
Throughout this paper we adopt a $\Lambda$CDM cosmology with Hubble constant $H_{0} = 70$
km\,s$^{-1}$ Mpc$^{-1}$, total dark matter density $\Omega_{M}=0.3$ and dark energy
density $\Omega_{\Lambda}=0.7$ in our calculations. 
Finally, in Section \ref{sec:discussion} we discuss and summarize our results.

\section{Observations and Data Reduction}
\label{sec:observations}

SN\,2002gh was discovered by the Nearby Supernova Factory \citep{aldering02} on 2002 October 5
(MJD=52552.41) at an unfiltered magnitude of 19.3 \citep{2002IAUC.7990....1W}. Previous to the
SN discovery, a non-detection on 2001 December 14 reported by \citet{2002IAUC.7990....1W}
places a 5$\sigma$ upper limit on the SN magnitude of 21.2 mag. The Nearby Supernova Factory
photometry of SN\,2002gh was calibrated to the USNO A1.0 ``red'' catalog, and therefore
is similar to $R$-band photometry (Wood-Vasey, private communication).

\begin{figure}
\centering
\includegraphics[width=90mm]{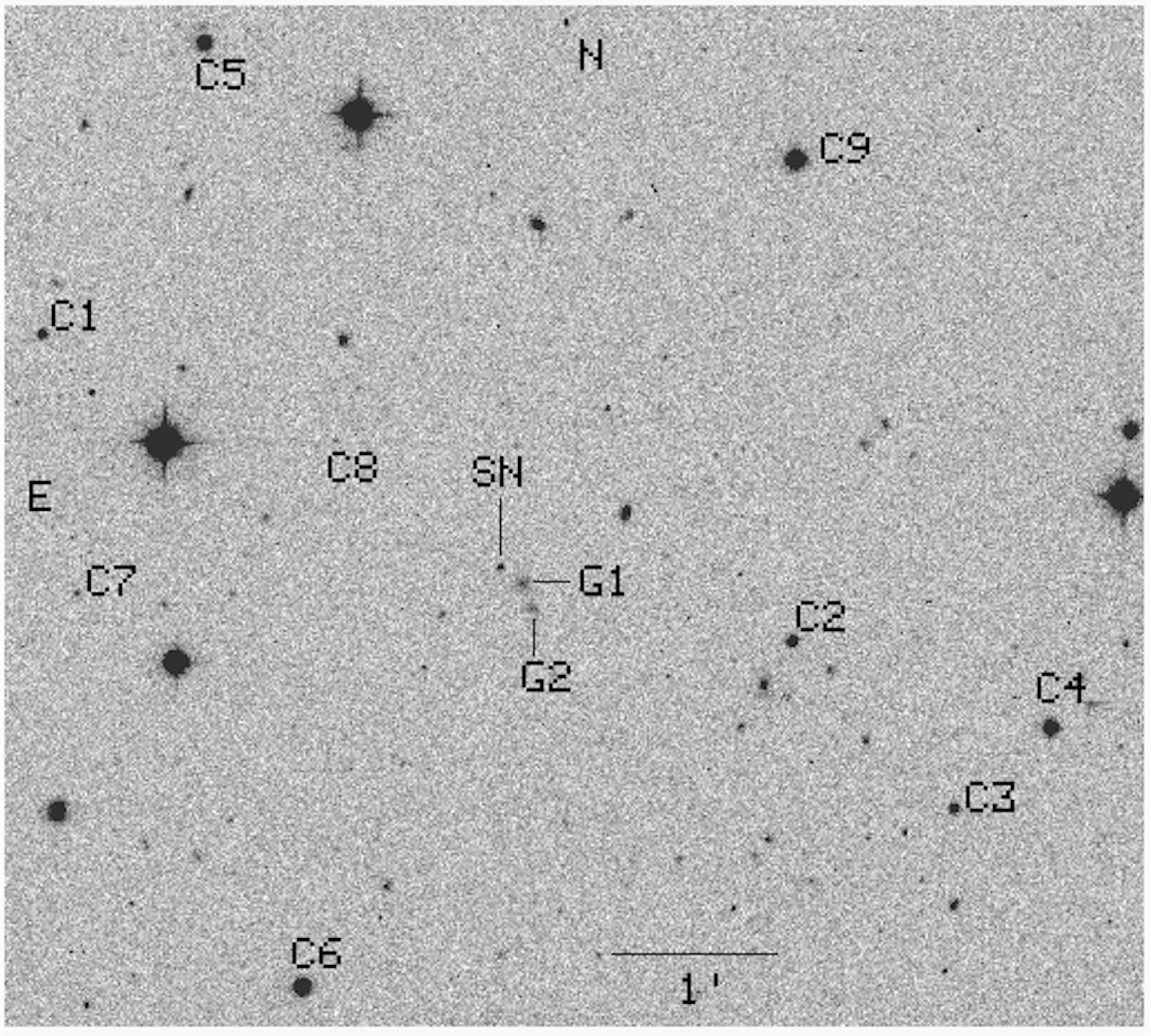}
\caption{Field of SN\,2002gh observed with the Swope 1 m telescope in the $V$-band when the
SN was near maximum light. North is up and east is to the left. We label SN\,2002gh along with nine comparison stars
used to derive the differential photometry of the SN. Two nearby galaxies to the south-west
of the SN are indicated as G1 and G2. Galaxy G1 has a redshift of 0.133 and was originally
identified as the possible host. The image scale is shown with a horizontal line near the bottom.}
\label{sn2002gh_fc_fig}
\end{figure}

\begin{figure}
\centering
\includegraphics[width=90mm]{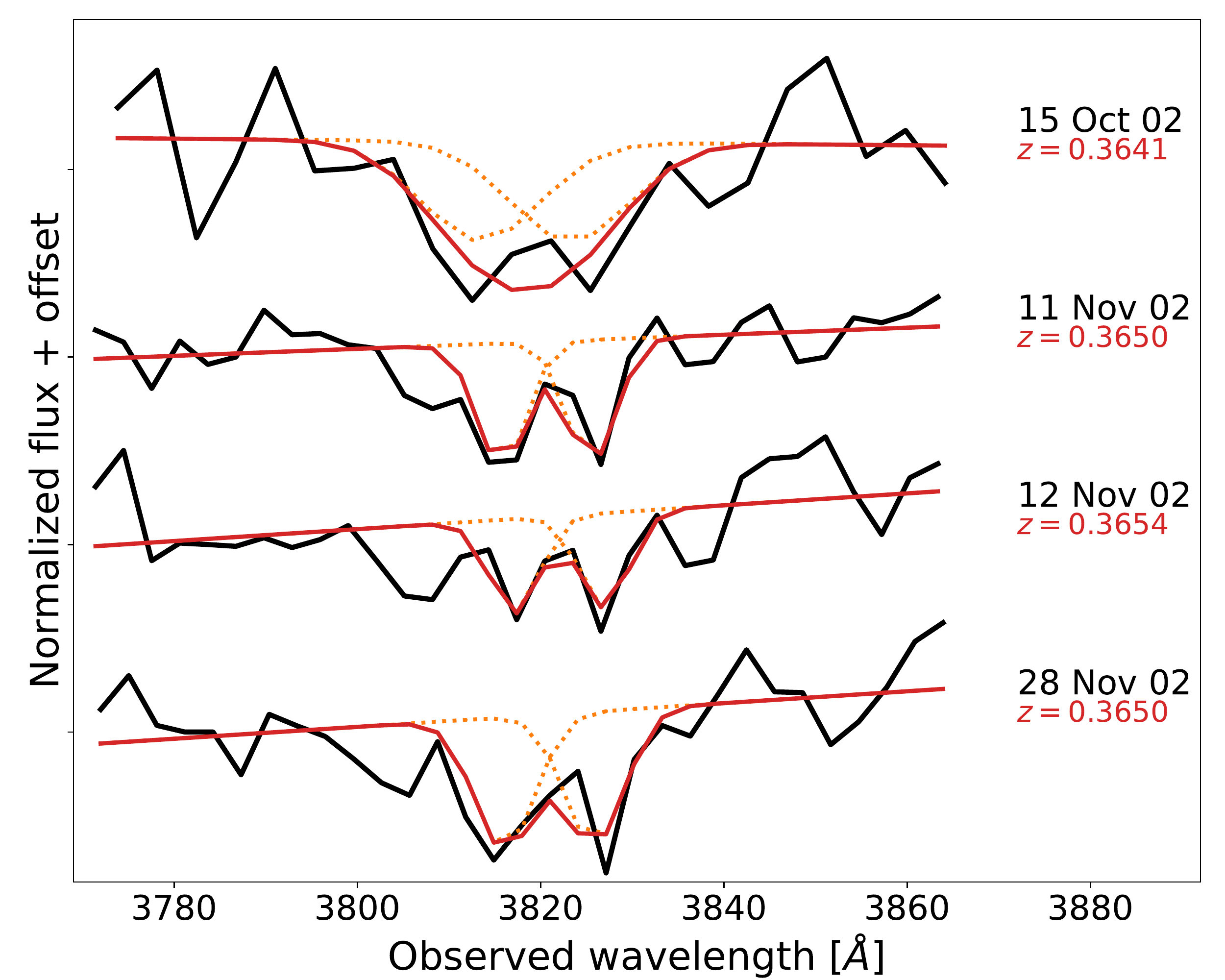}
\caption{Spectra of SN\,2002gh around the region of the \ion{Mg}{ii} $\lambda \lambda$\,$2796, 2803$ doublet
in the observer frame (in black), these spectra have been chosen because of their good signal-to-noise ratio
in this region. The epoch of the spectra is indicated on the right and the instruments and their characteristics
are listed in Table \ref{spec_summary_tab}.
The full Gaussian profile fits to the \ion{Mg}{ii} $\lambda \lambda$\,$2796, 2803$
doublet are shown in red and the individual Gaussian profiles are shown as orange-dotted lines.
The redshift obtained for each spectrum is given on the right, and the median
redshift for SN\,2002gh is $z=0.365 \pm 0.001$.}
\label{sn2002gh_mgii_fig}
\end{figure}

The SN was located at $\alpha$=03:05:29.46 $\delta$=$-$05:21:56.99 (J2000.0). No host galaxy was
clearly associated with SN\,2002gh in the CATS images to the limiting magnitude of the
deepest CATS images. The limiting magnitude of CATS images are $\sim 23$ mag in $BV$
and $\sim 21.5$ mag in $I$ band, this is 0.5 to 1 mag fainter than the faintest detection of
SN\,2002gh. A nearby galaxy at 8\farcs6 west and 6\farcs2 south from the SN  and at $z=0.133$, indicated as G1
in Fig. \ref{sn2002gh_fc_fig}, was originally identified as the possible host.
However, after inspecting our highest signal-to-noise ratio ($S/N$) spectra, we detected
\ion{Mg}{ii} $\lambda \lambda$\,$2796, 2803$ doublet narrow absorption lines from the host galaxy. 
We model the \ion{Mg}{ii} $\lambda \lambda$\,$2796, 2803$ doublet in these good $S/N$ spectra 
by fitting simultaneously two Gaussian profiles (see Fig. \ref{sn2002gh_mgii_fig}), where 
the relative central wavelengths are fixed accordingly and the variance of the Gaussian profiles
were also fixed according to the spectral resolution of the instrument used to obtain the spectra.
We adopt the median value from these four independent measurements and use three significant
digits for the uncertainty. This is to take into account any unaccounted uncertainty affecting
our redshift measurements, adopting $z = 0.365 \pm 0.001$ as the redshift for SN\,2002gh.
This redshift makes the SN spectra consistent with other hydrogen-poor SLSNe-I as discussed below
in Section \ref{sec:spec_analysis}.

The data were reduced using custom {\sc IRAF}\footnote{Image Reduction Analysis Facility,
distributed by the National Optical Astronomy Observatories (NOAO), which is operated by AURA Inc.,
under cooperative agreement with NSF.} scripts. A full description of the instruments
characteristics and the reduction procedures can be found in \citet{hamuy06, hamuy09}
and \citet{cartier14}.

\subsection{Photometry}
\label{photometry_obs_sec}

\begin{figure*}
\centering
\includegraphics[width=150mm]{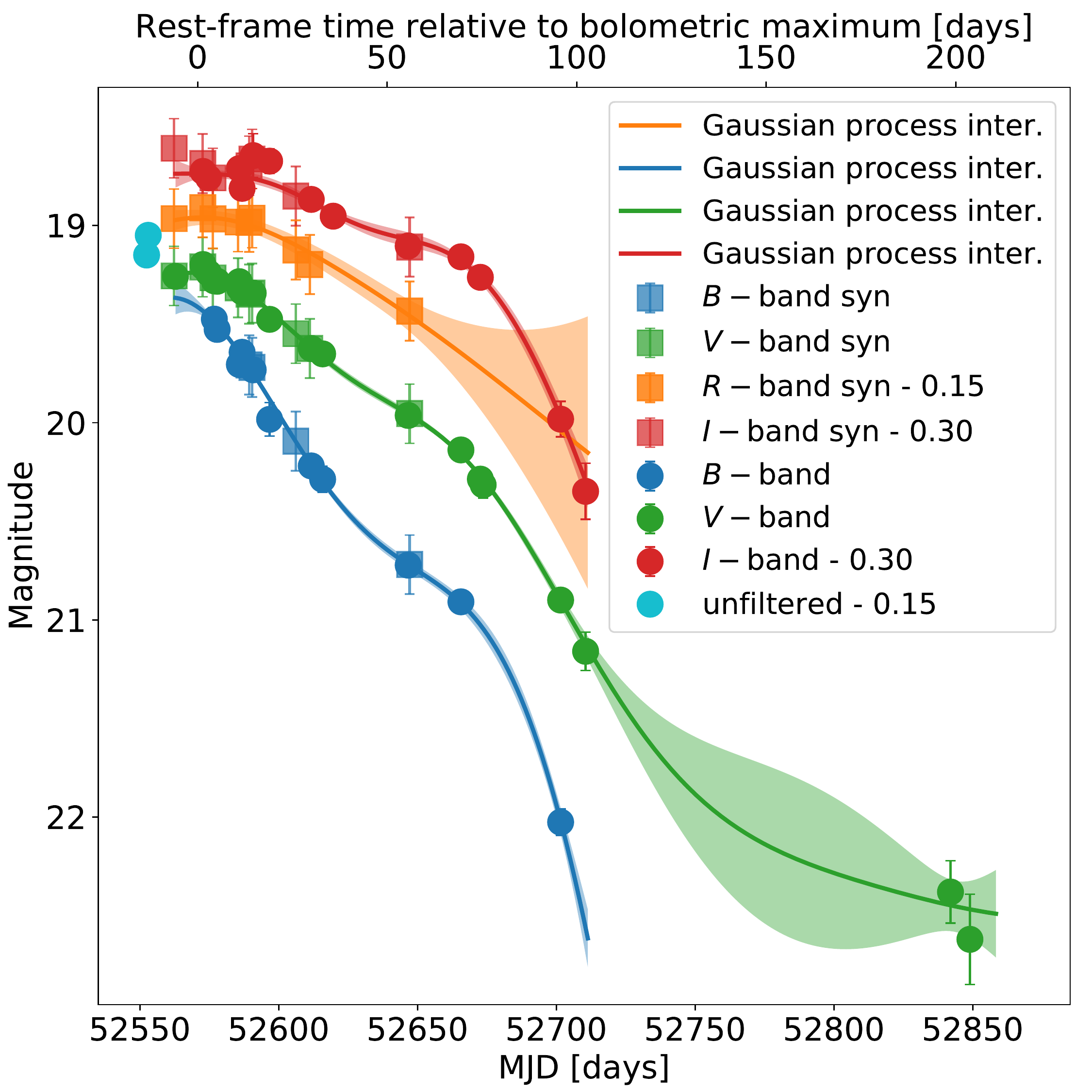}
\caption{Optical light curves of SN\,2002gh. We present $BVI$ photometry obtained
in the course of the CATS project \citep{hamuy09}, unfiltered photometry reported by
\citet{2002IAUC.7990....1W}, and $BVRI$ synthetic photometry computed from
the spectra after matching them to the interpolated photometry. The legend
is presented on the top right of the figure.}
\label{sn2002gh_lc_fig}
\end{figure*}

We obtained 22 epochs of $BVI$ optical photometry as part of the CATS project \citep{hamuy09},
beginning 10\,days after the SN discovery. Our first photometric observations were obtained six days before
bolometric maximum light and extend to $+204$ rest-frame days after peak. The photometry is summarized in Table
\ref{op_phot_tab} and the SN light curves are shown in Fig. \ref{sn2002gh_lc_fig}. The SN photometry was calibrated
differentially relative to a set of 9 stars in the field as shown in Fig. \ref{sn2002gh_fc_fig}.  
This local photometric sequence was calibrated in the Vega system on photometric nights with the Swope telescope at Las
Campanas Observatory. The instrumental magnitudes of the local standard stars and of the SN were converted to the standard
system using the colour terms and equations of \citet{hamuy09}. Their average $BVRI$ magnitudes are
summarized in Table \ref{loc_seq_tab}.

We interpolated the SN light curves using the {\sc python} Gaussian process module implemented in
{\sc scikit-learn} \citep{scikit-learn}. We used the radial-basis funnction (RBF) kernel for 
well-sampled light curves, while the Matern kernel with $\nu= 3/2$ was used for light curves
with long gaps, such as the $V$ band, or to extrapolate the $R-$band light curve. 
The kernel hyperparameters were optimized by maximizing the log-marginal-likelihood (LML).
As the LML may have multiple local optima, the optimizer was started 10 times. The first run was 
conducted starting from guess hyperparameter values. The subsequent runs were conducted
from hyperparameter values that have been chosen randomly from the range of allowed values. We present 
our interpolated light curves in Fig. \ref{sn2002gh_lc_fig}.

We used the interpolated light curves or the photometry from Table \ref{op_phot_tab} to
improve the flux calibration of the spectra presented in Section \ref{spectra_sec}. When photomometry was
obtained on the same night of a spectrum we used the observed photometry, otherwise we used the interpolated light curves.
The {\it mangling} of the spectra was perfomed using a low-order polynomial to match the spectral flux to
the SN photometry. The aim is to scale the flux calibration and correct any wavelength dependent slit
loss, and not to significantly alter the spectra. The application of this procedure to the 
spectra yields a small change in the spectral shape, if any.

Then, we obtained additional photometry by computing synthetic photometry from the {\it mangled} SN spectra
(see Fig. \ref{sn2002gh_lc_fig}), thus adding relevant photometric information in the $R$ band. 
The sythentic photometry was computed as described in \citet{bessell05}, using the Johnson-Cousins $BVRI$
response functions of \citet{bessell90}. The difference between the synthetic photometry and the interpolated
and observed photometry is always below 5\%, but we conservatively quote an uncertainty of 0.15 mag for the computed
synthetic photometry.

\begin{table*}
\caption{Optical photometry of SN\,2002gh}
\label{op_phot_tab}
\begin{tabular}{lccccc}
   \hline
   \multicolumn{1}{l}{Date UT} &
   \multicolumn{1}{c}{MJD} &
   \multicolumn{1}{c}{$B$}  &
   \multicolumn{1}{c}{$V$} &
   \multicolumn{1}{c}{$I$} &
   \multicolumn{1}{c}{Tel.} \\
   \hline
2002-10-15 & 52562.8 & --              & 19.258($0.010$) & --              & LDSS-2/Baade \\
2002-10-25 & 52572.7 & --              & 19.198($0.022$) & 19.026($0.028$) & CCD/Swope \\
2002-10-25 & 52572.8 & --              & 19.209($0.016$) & --              & LDSS-2/Baade \\
2002-10-27 & 52574.7 & --              & 19.241($0.015$) & 19.059($0.036$) & CCD/CTIO 0.9\,m \\
2002-10-29 & 52576.8 & 19.475($0.017$) & 19.281($0.014$) & --              & LDSS-2/Baade \\
2002-10-30 & 52577.8 & 19.527($0.020$) & 19.283($0.014$) & --              & WFCCD/du\,Pont \\
2002-11-07 & 52585.8 & --              & 19.316($0.024$) & --              & WFCCD/du\,Pont \\
2002-11-07 & 52585.8 & 19.705($0.031$) & 19.285($0.018$) & 19.012($0.054$) & CCD/Swope \\
2002-11-08 & 52586.8 & 19.643($0.016$) & 19.336($0.014$) & 19.112($0.038$) & CCD/Swope \\
2002-11-12 & 52590.8 & 19.732($0.039$) & 19.343($0.032$) & 18.946($0.111$) & CCD/Swope \\
2002-11-18 & 52596.7 & 19.983($0.085$) & 19.476($0.044$) & 18.974($0.061$) & CCD/Swope \\
2002-12-03 & 52611.7 & 20.219($0.019$) & 19.624($0.023$) & 19.168($0.037$) & WFCCD/du\,Pont \\
2002-12-07 & 52615.8 & 20.287($0.066$) & 19.651($0.055$) & --              & CCD/Swope \\
2002-12-11 & 52619.6 & --              & --              & 19.253($0.026$) & CCD/Swope \\
2003-01-07 & 52646.6 & 20.722($0.033$) & 19.963($0.020$) & 19.404($0.053$) & CCD/Swope \\
2003-01-26 & 52665.6 & 20.908($0.037$) & 20.138($0.021$) & 19.459($0.035$) & CCD/Swope \\
2003-02-02 & 52672.6 & --              & 20.286($0.024$) & 19.563($0.039$) & CCD/Swope \\
2003-02-03 & 52673.6 & --              & 20.315($0.068$) & --              & WFCCD/du\,Pont \\
2003-03-03 & 52701.5 & 22.026($0.067$) & 20.899($0.046$) & 20.282($0.090$) & WFCCD/du\,Pont \\
2003-03-12 & 52710.5 & --              & 21.159($0.097$) & 20.648($0.142$) & WFCCD/du\,Pont \\
2003-07-21 & 52841.9 & --              & 22.379($0.158$) & --              & CCD/Swope \\
2003-07-28 & 52848.9 & --              & 22.619($0.229$) & --              & CCD/Swope \\
\hline
\end{tabular}
\begin{tablenotes}
\item Numbers in parentheses correspond to 1\,$\sigma$ statistical uncertainties.
\end{tablenotes}
\end{table*}

\begin{table*}
\caption{$BVRI$ photometric sequence around SN\,2002gh.}
\label{loc_seq_tab}
\begin{tabular}{lcccccc}
   \hline
   \multicolumn{1}{l}{Star} &
   \multicolumn{1}{l}{R.A.} &
   \multicolumn{1}{c}{Dec}  &
   \multicolumn{1}{c}{$B$}  &
   \multicolumn{1}{c}{$V$}  &
   \multicolumn{1}{c}{$R$}  &
   \multicolumn{1}{c}{$I$}  \\
   \hline
C1 & 03:05:40.66 & $-$05:20:36.77 & 18.997(0.040) & 18.104(0.013) & 18.450(0.017) & 17.042(0.009) \\
C2 & 03:05:22.31 & $-$05:22:21.09 & 18.466(0.042) & 17.572(0.015) & 17.936(0.015) & 16.529(0.009) \\
C3 & 03:05:18.29 & $-$05:23:20.11 & 18.876(0.019) & 17.799(0.025) & 18.106(0.015) & 16.425(0.015) \\
C4 & 03:05:16.00 & $-$05:22:50.10 & 16.684(0.015) & 15.955(0.019) & --            & 15.029(0.010) \\
C5 & 03:05:36.90 & $-$05:18:49.67 & 17.067(0.037) & 16.067(0.015) & 16.399(0.015) & 15.005(0.008) \\
C6 & 03:05:33.99 & $-$05:24:30.93 & 16.094(0.031) & 15.429(0.024) & 15.987(0.015) & 14.656(0.010) \\
C7 & 03:05:39.72 & $-$05:22:11.62 & 22.764(0.538) & 21.385(0.093) & 21.892(0.366) & 18.795(0.026) \\
C8 & 03:05:33.46 & $-$05:21:12.59 & 22.869(0.339) & 21.296(0.162) & 20.873(0.146) & 18.665(0.024) \\
C9 & 03:05:22.52 & $-$05:19:26.92 & 15.590(0.031) & 14.903(0.014) & 15.416(0.015) & --            \\
\hline
\end{tabular}
\begin{tablenotes}
\item Numbers in parentheses correspond to 1\,$\sigma$ statistical uncertainties.
\end{tablenotes}
\end{table*}

\subsection{Spectroscopy}
\label{spectra_sec}

\begin{figure*}
\centering
\includegraphics[width=176mm]{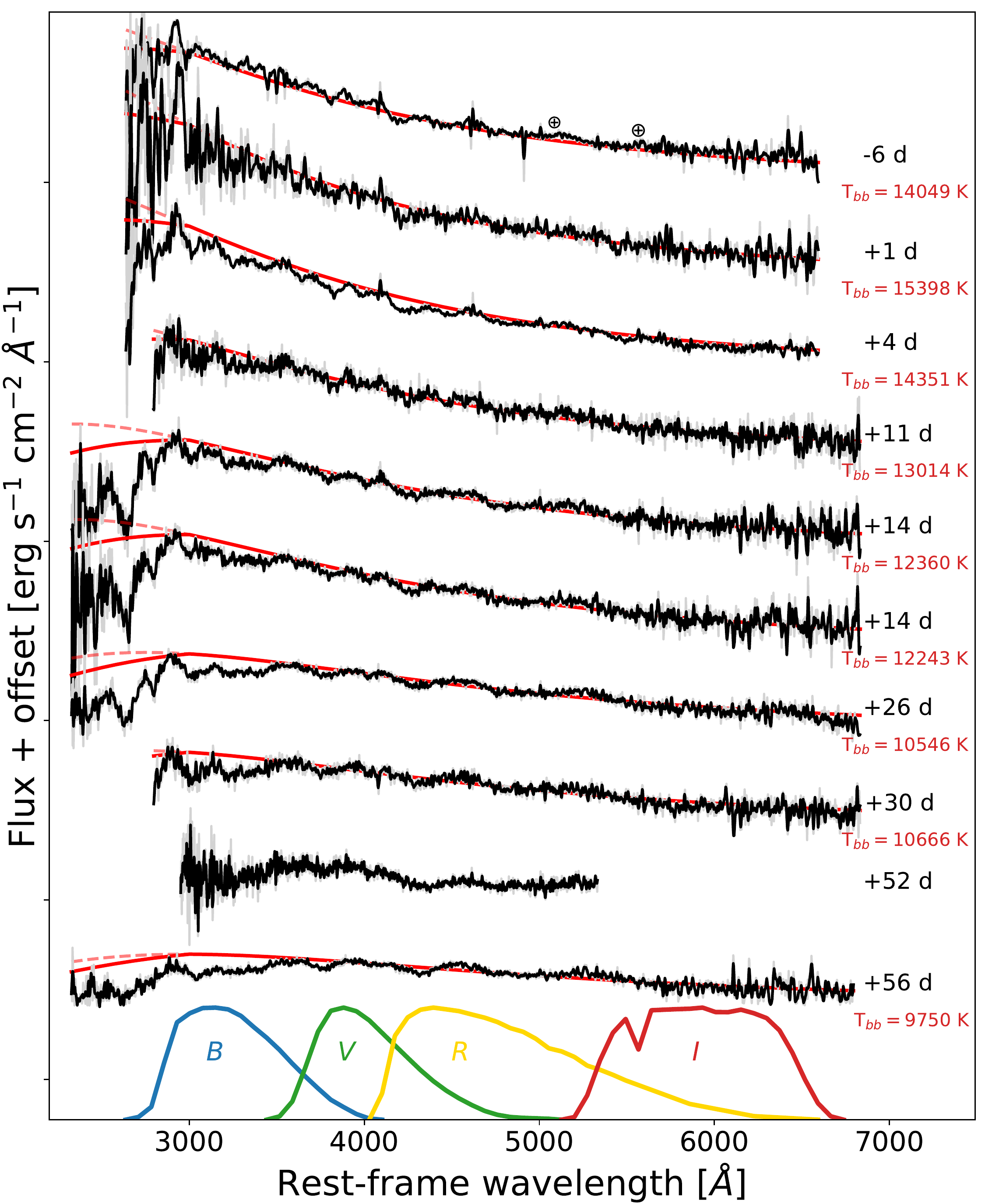}
\caption{Spectroscopic sequence of SN\,2002gh. The spectra have been mangled in order to match
the $BVI$ interpolated photometry at their corresponding epoch, yielding accurate flux
calibration. Then we corrected the spectra by a Galactic reddening of $E(B-V) = 0.0593$\,mag \citep{schlafly11}
in the line-of-sight using the \citet{cardelli89} reddening law with
$R_{V} = 3.1$. The spectra are corrected by the host galaxy redshift to place the spectra
in the rest-frame system. The red solid lines correspond to modified blackbody fits to the SN spectra
with a linear flux suppression below the ``cutoff'' wavelength at 3000 \AA, and the dashed red lines
correspond to the normal blackbody without flux suppression. On the right we give the phase relative
to the estimated rest-frame time since bolometric maximum and the blackbody temperature ($T_{\mathrm{bb}}$).
At the bottom of the figure we show the $BVRI$ normalized transmissions at the SN rest-frame.
}
\label{sn2002gh_spec_fig}
\end{figure*}

A total of ten spectra of SN\,2002gh were obtained, spanning from -6 to +56
rest-frame days relative to the estimated bolometric maximum. We summarize spectroscopic
observations in Table \ref{spec_summary_tab}, and the spectral sequence is presented
in Fig. \ref{sn2002gh_spec_fig}. We improved the flux calibration by using interpolated
photometry (see Section \ref{photometry_obs_sec}) and we fitted a modified blackbody
model to the spectral sequence to estimate the blackbody radius ($R_{bb}$) and temperature ($T_{bb}$).
To fit the modified blackbody model we selected pseudo-continuum emission regions.
For this end, we use regions of the spectra redwards of $2900$ \AA\ \citep[see e.g.,][]{prajs17}
and devoid of any strong spectral features.

\begin{table*}
 \caption{Summary of Spectroscopic Observations of SN~2002gh.}
 \label{spec_summary_tab}
 \begin{tabular}{@{}lcccccc}
  \hline
  Date UT & MJD        & Phase  & Instrument/ & Wavelength  & Dispersion  & Resolution  \\
          &            & (days) & Telescope   & Range (\AA) & (\AA/pixel) & (\AA)       \\
  \hline
  2002-10-15 & $52562.3$ & $-6.2$  & LDSS-2/Baade        & $3600 - 9000$  & $4.3$ & $13.5$ \\
  2002-10-25 & $52572.6$ & $+1.3$  & LDSS-2/Baade        & $3600 - 9000$  & $4.3$ & $13.5$ \\
  2002-10-29 & $52576.3$ & $+4.0$  & LDSS-2/Baade        & $3600 - 9000$  & $4.3$ & $13.5$ \\
  2002-11-07 & $52585.3$ & $+10.6$  & WFCCD/du\,Pont      & $3820 - 9330$  & $3$   &  $6$ \\
  2002-11-11 & $52589.3$ & $+13.6$ & B\&C Spec./Baade    & $3180 - 9315$  & $3$   &  $6$ \\
  2002-11-12 & $52590.3$ & $+14.3$ & B\&C Spec./Baade    & $3180 - 9315$  & $3$   &  $6$ \\
  2002-11-28 & $52606.2$ & $+25.9$ & B\&C Spec./Baade    & $3180 - 9312$  & $3$   &  $6$ \\
  2002-12-03 & $52611.2$ & $+29.6$ & WFCCD/du\,Pont      & $3820 - 9330$  & $3$   &  $6$ \\
  2003-01-03 & $52642.2$ & $+52.3$ & Mod. Spec./du\,Pont & $3780 - 7280$  & $2$   &  $7$ \\
  2003-01-08 & $52647.1$ & $+55.9$ & B\&C Spec./Baade    & $3180 - 9280$  & $3$   &  $6$ \\
  \hline
 \end{tabular}
\end{table*}

\section{Analysis}
\label{sec:analysis}

\subsection{Line identification and spectral comparison}
\label{sec:spec_analysis}

The spectral sequence of SN\,2002gh is presented in Fig. \ref{sn2002gh_spec_fig}, the SN evolves smoothly from a very
blue nearly featureless spectrum to a redder spectrum dominated by moderate absorption lines.
We divide, arbitrarily, the spectral evolution of SN\,2002gh in four characteristic phases, namely, an early phase,
followed by a cooling phase, a post-maximum phase, and a cool-photospheric phase. To study the spectral
evolution we compare the SN\,2002gh spectra with spectra of SNe from the literature. The SNe used for comparison
satisfy the following conditions: 1) have been well studied in the literature, thus we can use previous identifications to
guide our line identification, and 2) show spectral features similar to SN\,2002gh. These SNe are not necessarily representative of
the whole diversity of the SLSN spectra \citep[see e.g., Fig. 4 of][]{anderson18,quimby18}. The literature spectra
were obtained from the WISeREP repository \citep{yaron12}.

To distinguish the main ion(s) that contribute to form a spectral feature we follow a similar approach to
\citet{galyam19b}. First, all permitted lines of 11 ions (including \ion{He}{i}) are considered
and compiled from the National Institute of Standards and Technology (NIST), covering the wavelength range from 2500 to 7000 \AA.
The relative intensities of the lines for a ion are normalized by the maximum relative intensity of that ion in this
wavelength range. Then only lines with normalized relative intensities greater than 0.5 are considered for the line identification,
but a few exceptions are made for some lines. 
In a few cases there are strong UV lines, with very large relative intensities, making it impractical to apply the
normalization criteria described previously. The application of this criteria would not consider some optical lines
that can be distinguished in the spectra, or would leave out a few lines that have been repeatedly associated with
a spectral feature in the literature. Hence, we decided to keep these lines in our line list for completeness.

Sometimes multiple strong lines of an ion are located close in wavelength (\textless 100 \AA), and due to the large ejecta
expansion velocities (thousands of km s$^{-1}$), these lines appear blended contributing to form a single spectral feature.
To simplify the line identification in these cases, they are represented as a single line having a mean wavelength computed
using their relative intensities as weights for the mean. The weighted means are computed over wavelength
intervals of 30 to 100 \AA. An example of this methodology is the H\&K \ion{Ca}{ii} doublet $\lambda \lambda \, 3933.66, 3968.47$,
which in the SN spectra yield a blended feature with a weighted mean wavelength of 3950.29 \AA.

With these ion line lists in hand, and in combination with line identification and ratiative transfer models of SLSNe from
the literature \citep[e.g.,][]{mazzali16,quimby18,dessart19,galyam19b} we associate these ion lines to spectral features.
We also considered \citet{hatano99} synthetic spectra computation for C/O-rich and carbon-burned compositions
to guide our line identification.

\subsubsection{Early phase and maximum ($t \lessapprox +5$ d)}

\begin{figure*}
\centering
\includegraphics[width=150mm]{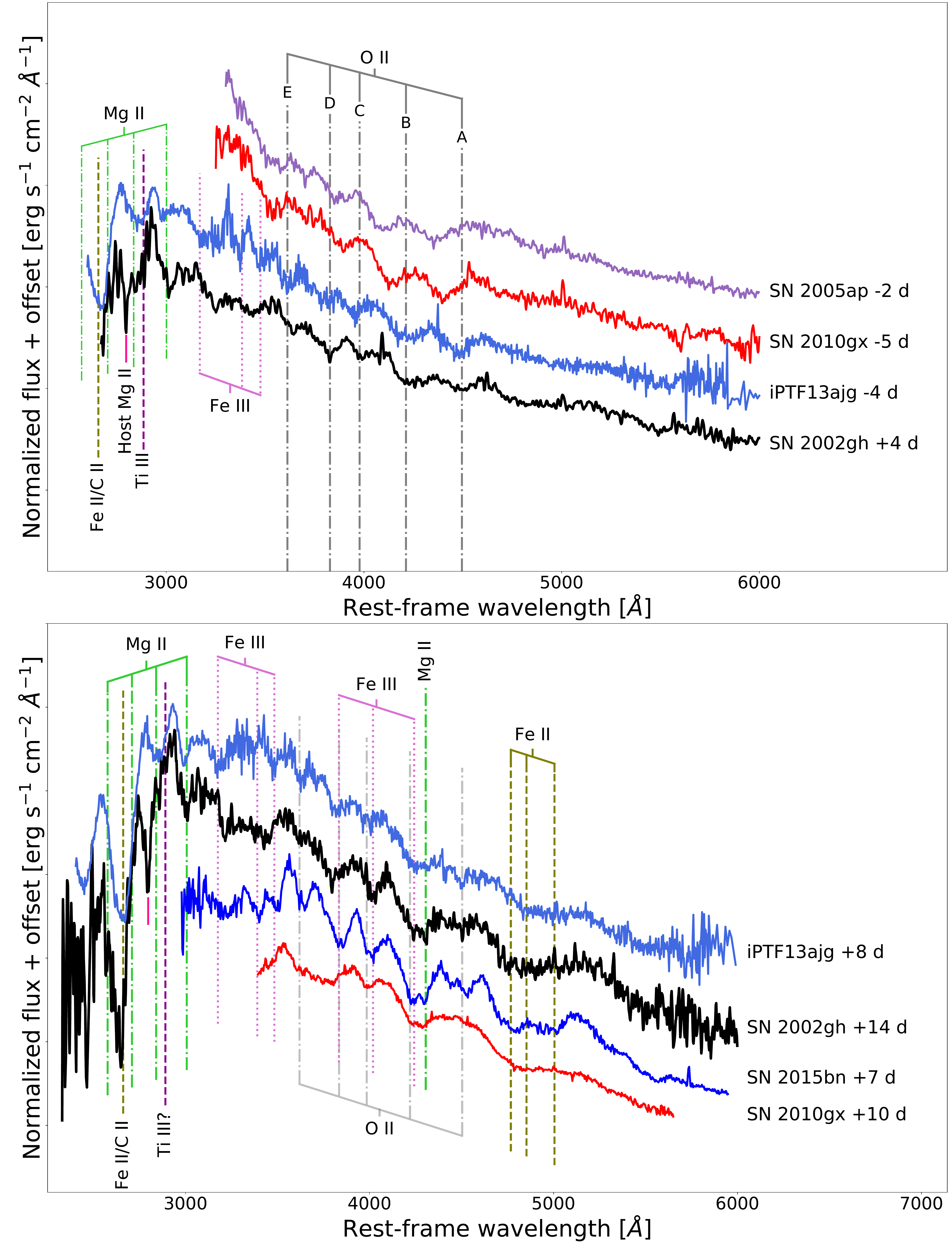}
\caption{Comparison between the early phase spectra of SN\,2002gh, SN\,2005ap \citep{quimby07}, SN\,2010gx \citep{pastorello10},
iPTF13ajg \citep{vreeswijk14}, and LSQ14bdq \citep{nicholl15} (top panel), these SNe show the characteristic \ion{O}{ii} spectral features in the
3500-5000 \AA\ region. The vertical dot-dashed lines represent the effective wavelengths for the A, B, C, D, and E \ion{O}{ii} spectral features
identified by \citet{quimby18} at an expansion velocity of $10,000$ \kms. We identify and mark other ions with vertical
lines such as the \ion{Ti}{iii} $\lambda \, 2984.747$ line identified by \citet{mazzali16}, \ion{C}{ii}, \ion{Mg}{ii}, \ion{Fe}{ii} and \ion{Fe}{iii}
lines at the same expansion velocity of $10,000$ \kms. We mark the host galaxy \ion{Mg}{ii} $\lambda \lambda$\,$2796, 2803$ doublet
absorption feature with a pink tick. On the bottom panel, we present the comparison between the post-maximum cooling phase spectra of SN\,2002gh,
SN\,2010gx, iPTF13ajg, and SN\,2015bn \citep{nicholl16}. In addition to the correction by the host galaxy recession velocity, the spectrum
of SN\,2015bn has been shifted to the blue by 2500 \kms\ to match the SN\,2002gh spectrum at this phase. As in the top panel,
the lines of several ions are marked and identified with vertical lines at an expansion velocity of 9500 \kms.
}
\label{spec_comp_maxpremax_fig}
\end{figure*}

The spectra of SLSNe-I near to maximum light are characterized by a blue continuum which is well described by a blackbody, and
by the presence of the distinctive \ion{O}{ii} lines in the wavelength range from 3500 to 5000\,\AA\ 
\citep[see][]{quimby18}. The \ion{O}{ii} features are the hallmark of hydrogen-poor SLSNe before and close to maximum light,
and these lines are extremely rare in other CCSNe as it requires oxygen to be excited to a high energy level
\citep{mazzali16}. The spectra of SN\,2002gh from -6 to +4\,days display these distinctive signatures 
(see top panel of Fig. \ref{spec_comp_maxpremax_fig}). 

As shown in the top panel of Fig. \ref{spec_comp_maxpremax_fig}, the expansion velocity of SN\,2002gh measured from
the minimum of the \ion{O}{ii} is about $10,000$ \kms\ and its photospheric temperature is about $14,000$\,K. The strong spectral
feature at $2830$ \AA\ has been associated with \ion{Ti}{iii} in iPTF13ajg by \citet{mazzali16}. In Fig. \ref{spec_comp_maxpremax_fig}
we confirm the potential contribution of \ion{Ti}{iii}, and we add the very likely contribution from \ion{Mg}{ii}
to this feature, both ions at an expansion velocity of about $10,000$ \kms. In the spectra of SN\,2002gh the blue side of this feature
is affected by the \ion{Mg}{ii} $\lambda \lambda$\,$2796, 2803$ doublet absorption from the host galaxy, which is indicated with a
pink tick mark in Fig. \ref{spec_comp_maxpremax_fig}. After a careful inspection, the minimum of this absorption feature seems to be
redwards of the host galaxy absorption, and favours \ion{Mg}{ii} as the dominant ion for this feature in SN\,2002gh.

We identify three absorption features in LSQ14bdq as tentative \ion{Fe}{iii} lines, in the region between $3100$ \AA\ and $3500$ \AA.
We further identify these features in the early spectra of iPTF13ajg (see also the right-panel of Fig. \ref{spec_comp_maxpremax_fig}),
but these features are not clearly present in SN\,2002gh. We note that other strong \ion{Fe}{iii} lines have wavelengths coincident
with \ion{O}{ii} features (see Sec. \ref{cooling_phase_sec} below), furthermore the \ion{C}{iii} $\lambda \, 4649.97$ feature is
coincident and may contribute to the \ion{O}{ii} A feature at this phase. The early phase spectra of SN\,2002gh show similarities
with the early spectra of LSQ14bdq \citep{nicholl15}, and with the early and near maximum spectra of iPTF13ajg \citep{vreeswijk14}.

\subsubsection{Cooling phase ($ +4$ d $\textless t \lessapprox +15$ d)}
\label{cooling_phase_sec}

The cooling begins at a few days after maximum and ends between 10 and 20\,days past maximum. It is characterized by a decrease in the strength of
the characteristic \ion{O}{ii} lines, until they disappear at the end of this phase. This is a consequence of the decrease in the ionization
state of the SN ejecta after maximum light. However, the SN ejecta remains hot enough for \ion{Fe}{iii} features to take the place
of \ion{O}{ii} features, while the latter decrease their strength significantly. This evolution is reflected in: 1) \ion{O}{ii} A and E features
fade quicker than other \ion{O}{ii} features after maximum, 2) the \ion{O}{ii} C and D features, which are coincident with \ion{Fe}{iii} lines,
remain visible in the SN spectra for longer time than A and E features, and 3) in the meantime, the \ion{O}{ii} B feature which is blended
with \ion{Fe}{iii}, \ion{Fe}{ii} and \ion{Mg}{ii} lines, increases its strength and quickly shifts to the red. At this phase,
the \ion{O}{ii} B feature exhibits a distinctive double absorption due to \ion{Fe}{iii} in the blue and to \ion{Mg}{ii} in the red.
In the bottom panel of Fig. \ref{spec_comp_maxpremax_fig} we illustrate the \ion{Fe}{iii} lines and the \ion{Mg}{ii} $\lambda \, 4446.31$ feature
at an expansion velocity of $9500$ \kms\ in the region previously dominated by \ion{O}{ii} lines. These two minima can be clearly distinguished
in SN\,2015bn and SN\,2010gx. Contemporary to this, \ion{Fe}{ii} lines give shape to the broad spectral feature in the region between 4500-5250 \AA\
(see bottom panel of Fig. \ref{spec_comp_maxpremax_fig}).

As before, the \ion{Fe}{iii} absorption features are tentatively distinguished in the spectra of iPTF13ajg and SN\,2015bn in the region between
$3100$ \AA\ and $3500$. However, the identification of these features is less clear in SN\,2002gh. We associate the feature at
3000 \AA\ with \ion{Mg}{ii} in SN\,2002gh and in iPTF13ajg. The spectral feature at $2830$ \AA\ is now shifted blueward in these SNe,
in a better agreement with the \ion{Mg}{ii} line rather than with the \ion{Ti}{iii} line, suggesting a decrease in the contribution of the latter
ion to this feature. We associate the strong 2670 \AA\ feature with \ion{Fe}{ii} and \ion{Mg}{ii} lines, and note that the minimum of
this feature is bluer in SN\,2002gh. Possible explanations for this are; 1) a higher expansion velocity of the line producing this feature
in SN\,2002gh, 2) a contribution from an unidentified ion not present in the iPTF13ajg spectrum or 3) host galaxy absorption lines
shifting the line to the blue as in the case of the \ion{Mg}{ii} host galaxy absorption line that {\it `shifts'} the 2830 \AA\ feature to the blue.
Several strong \ion{Fe}{ii} lines on the blue side of this feature at the host galaxy rest-frame can be potentially identified, providing
support to the latter possibility.

The spectra of SN\,2002gh from +9 to +12\,days are representative of the cooling phase. The average decrease of the SN\,2002gh ejecta
temperature is about 210\,K per day, with an approximate ejecta temperature within the range of about $13,500$ to $11,000$\,K in the course
of this phase. We find that the spectra of SN\,2002gh at this phase are similar to iPTF13ajg and SN\,2015bn (after shifting the latter
spectrum to the blue in Fig. \ref{spec_comp_maxpremax_fig} for a better comparison with the spectral features of SN\,2002gh).

\subsubsection{Post-maximum phase ($ +15$ d $\lessapprox t \lessapprox +40$ d)}

The beginning of the post-maximum phase can be defined as when high-ionization lines such as \ion{O}{ii} and \ion{Fe}{iii} lines vanish,
and the dominant spectral features are produced by \ion{Fe}{ii}, \ion{Mg}{ii} and other IME's. This happens at about
10 to 20 days after maximum. These changes make hydrogen-poor SLSNe exhibit spectral features similar to SNe Ic and Ic-BL
at an early phase (Fig. \ref{spec_comp_coolphotos_fig}), but the spectral features of SLSNe are shallower and have a bluer
continuum.

In Fig. \ref{spec_comp_coolphotos_fig} we distinguish the 3800 \AA\ feature produced by \ion{Mg}{ii} and the H\&K \ion{Ca}{ii}
lines, the 4315 \AA\ feature produced by \ion{Mg}{ii} and \ion{Fe}{ii} lines and the broad feature between 4500-5250 \AA\
produced mainly by \ion{Fe}{ii} lines, but with contributions from other IME's such as \ion{Mg}{ii}. Notably,
the 4315 \AA\ feature has evolved from being produced by the \ion{O}{ii} B feature, then produced by a combination of \ion{Fe}{iii},
\ion{Fe}{ii} and \ion{Mg}{ii} with a distinctive double absorption feature such as in SN\,2015bn and SN\,2010gx, to a single absorption produced
by \ion{Mg}{ii} and \ion{Fe}{ii} at this phase. Note that there is an \ion{Fe}{ii} line close in wavelength to the \ion{Mg}{ii} line
in this feature, which we do not mark in Fig. \ref{spec_comp_coolphotos_fig} to avoid crowding. Additional
\ion{Fe}{ii} lines bluewards and redwards in wavelength also shape this feature.

At UV wavelengths, the spectral features at $2670$ \AA\ and $2830$ \AA\ remain, and \ion{Fe}{ii} and \ion{Mg}{ii} are the
main suspects to explain these features. For first time the \ion{Ca}{ii} features are detected, and they contribute to the
spectral features  at 3080 \AA\ and 3800 \AA\ blended with \ion{Mg}{ii} lines (see Fig. \ref{spec_comp_coolphotos_fig}).
The spectra of SN\,2002gh at +24 and +28\,days are representative of this phase, where the ejecta expansion velocity is about
$9000$ \kms\ and $T_{\mathrm{bb}} \sim 10,500$\,K.

\subsubsection{Cool-photospheric phase ($t \textgreater +40$ d)}

\begin{figure*}
\centering
\includegraphics[width=150mm]{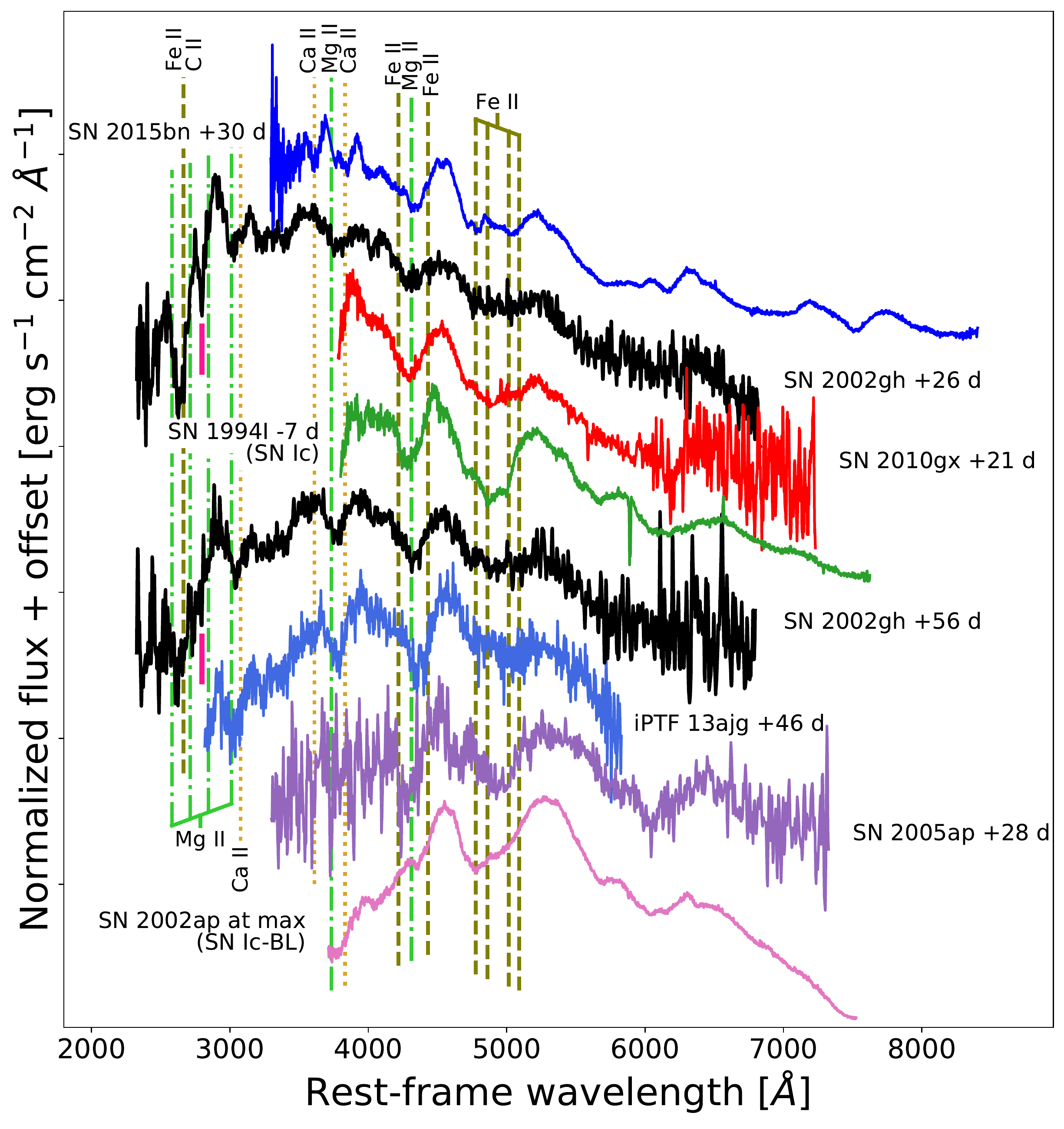}
\caption{Post-maximum and cool-photospheric phase comparison. We mark and identify the lines of several ions with
vertical lines at an expansion velocity of 9000 \kms. For comparison, we present the spectra of the Type Ic
SN\,1994I and the Type Ic-BL SN\,2002ap \citep{modjaz14}, and the  hydrogen-poor SLSN 2005ap, SN\,2010gx, iPTF13agj,
and SN\,2015bn. As in Fig. \ref{spec_comp_maxpremax_fig}, the spectrum of SN\,2015bn has been shifted to the blue by 2500 \kms.
As before, we highlight the \ion{Mg}{ii} $\lambda \lambda$\,$2796, 2803$ doublet host galaxy absorption in SN\,2002gh spectra
with a pink tick.
}
\label{spec_comp_coolphotos_fig}
\end{figure*}

As the ejecta temperature decreases the resemblance of SLSNe to SNe Ic and Ic-BL becomes more evident, the pseudo-continuum
becomes redder and the strengths of the lines increase, as can be seen in Fig. \ref{spec_comp_coolphotos_fig}. When the ejecta
temperature of SLSNe is about or below $10,000$\,K the strength of the 4315 \AA\ feature in SLSNe becomes similar to SNe Ic.
This is the case of SN\,2002gh from about +40\,days, when the ejecta temperature is about $9500$\,K and the ejecta expansion velocity is
slightly below $9000$ \kms.

In Fig. \ref{spec_comp_coolphotos_fig}, we notice that while the 4315 \AA\ feature in SLSNe has reached a strength roughly similar as
in the Ic SN\,1994I, the spectral feature between 4500-5250 \AA, although stronger than in previous phases, is shallower than
in SNe Ic and Ic-BL. The UV emission has decreased due to the decline of the ejecta temperature and due to an increase of absorption
from IME's and \ion{Fe}{ii} lines. The spectral features at $2670$ \AA\ and $2830$ \AA\ remain. The SLSN emission it is now slowly
evolving towards the pseudo-nebular phase.

\subsection{Host galaxy characterization}
\label{sec:host_galaxy}

The field of the SN explosion has been covered by several modern digital 
surveys over the past decades. With the aim of identifying and characterising the
SN\,2002gh host galaxy, we searched for host galaxy candidates close to the SN position in
these deep surveys. The nearest astronomical source to the SN site detected by both
SDSS and Pan-STARSS is galaxy G1 shown in Fig. \ref{sn2002gh_fc_fig}, and already discussed
in Section \ref{sec:observations}.

\begin{figure*}
\centering
\includegraphics[width=180mm]{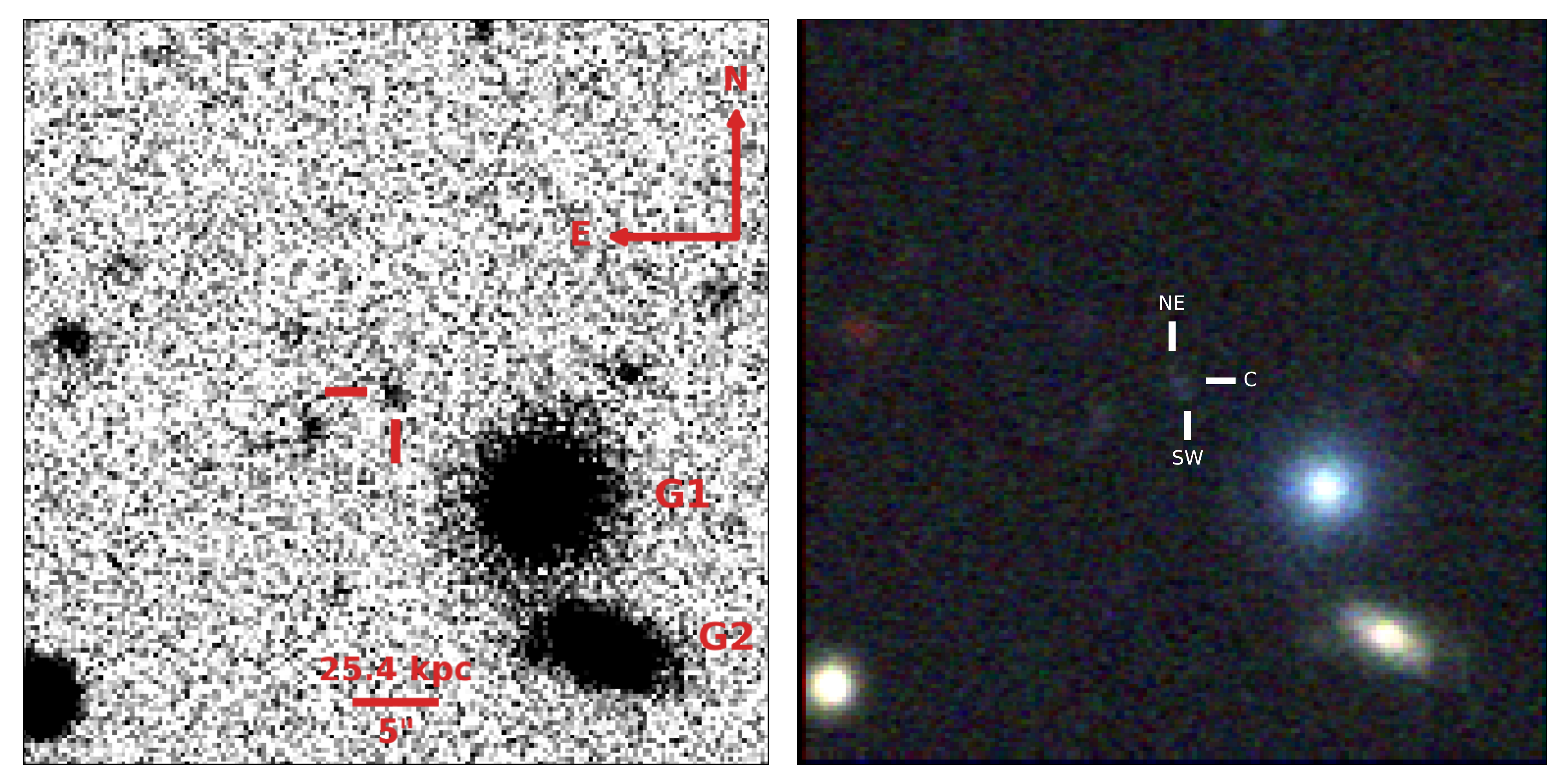}
\caption{In the left panel, we present a greyscale image centred on the position of SN\,2002gh,
resulting from stacking all the $griz$ DECam frames available. We indicate the host galaxy position with red tick marks, and we label
the foreground galaxies G1 and G2. In the right panel, we present a $grz$ colour image composite made of deep
image stacks in these filters. In the colour image, the position of the north-east, central and south-west faint distinct components 
detected in the $r$ band associated to potential host galaxy of SN\,2002gh, are indicated with a white tick marks.
The field-of-view of both images is 0\farcm72 $\times$ 0\farcm72, and at the bottom of the left panel we give as reference the projected
physical distance of 5\farcs0 at the SN redshift. 
}
\label{sn2002gh_host_gal_fig}
\end{figure*}

We queried the NOAO archive for DECam archival frames containing the SN region and
after inspecting them, we noticed some faint emission close to the SN site on some
of the individual DECam frames. However, the number of counts is too low to claim a detection
on these individual frames. Therefore, to sum all possible photons, we stacked all $griz$ frames
for each individual filter, and we also made a multi-band image stack
with all the $griz$ frames using {\sc SWarp} \citep{2002ASPC..281..228B}.
Figure \ref{sn2002gh_host_gal_fig} shows the result of the multi-band $griz$ image stack
and a colour composite image made of the individual $g$, $r$, $z$ stacks.

On the multi-band image stack, we detected an extended object at $\sim 10 \sigma$ flux level above the background
with centroid equatorial coordinates of $\alpha =$03:05:29.47 $\delta=-$05:21:57.08 (J2000) located 0\farcs13 south-east from the SN
location corresponding to a projected distance of 0.67 kpc, at the redshift of the SN. We did not detect this source
in the $i$-band image stack. In the $g$-band deep stack, we detected this source at $\sim 9 \sigma$ level. On
the $r$-band deep stack we detected three faint sources within $\sim$ 1\farcs5 of the SN position. The first
source is detected at $\sim 6 \sigma$ and the peak of the emission is 1\farcs01 or at a projected distance of
5.14 kpc north-east from the galaxy centroid measured in the multi-band image stack. The central source is
detected at $\sim 10 \sigma$ level and is located 0\farcs08 or at a projected distance of 0.44 kpc to the
north-west of the galaxy centroid, this is essentially coincident with the multi-band image centroid.
Finally, the third source is detected at $\sim 8 \sigma$ and the peak is 0\farcs96 or at a projected distance of
4.90 kpc to the south-west of the galaxy centroid. Although the latter source is not detected
as a different source in the $g$-band, it is clear that there are two emission peaks in the $g$-band image stack,
one coincident with the main central source and one coincident with the south-west source. The three sources
detected in the $r$-band can be distinguished in the colour composite image on Fig. \ref{sn2002gh_host_gal_fig}. In the $z$-band
we detected a source at the $\sim 6 \sigma$ level with the peak located 0\farcs36 or at a projected distance of 1.82 kpc to the
north-east from the multi-band galaxy centroid. Again this is essentially coincident with the main multi-band
image centroid. The $z$-band source is very weak and does not show evidence of structured emission. Although the small
projected distance between the host galaxy candidate reported here and the SN is a strong reason to think that this
is the SN host, further spectroscopic confirmation of the galaxy redshift is required for a solid association.   
The multi-peak morphology found in the SN\,2002gh's host candidate is common in hydrogen-poor SLSN host galaxies.
Analysing {\it Hubble Space Telescope} images, \citet{lunnan15} found that roughly half of the hydrogen-poor SLSN
host galaxies exhibit a morphology that is either asymmetric, off-centre or consisting of multiple peaks. Recently,
\citet{vanggaard20} found that hydrogren-poor SLSN host galaxies are often part of interacting systems, with
$\sim 50$\% having at least one major companion within 5\,kpc.

Assuming that the source detected is the SN host, we note that at this redshift the usual emission lines of SLSN-I host galaxies,
such as [\ion{O}{ii}] $\lambda \lambda$\,$3726,3729$ doublet emission lines lie wthin the $g$ band, the H$\beta$
and the [\ion{O}{iii}] $\lambda \lambda$\,$4959,5007$ doublet emission lines lie within the $r$ band,
and the H$\alpha$ and [\ion{S}{ii}] $\lambda \lambda$\,$6717,6731$ doublet emission lines lie within the $z$ band.
These emission lines are strong in regions with recent or ongoing star-formation, in particular in Extreme Emission
Line Galaxies \citep[EELGs; e.g.][]{cardamone09,atek11,amorin14,amorin15}, and therefore the peaks detected could be tracing regions of
strong star-formation within the irregular host or could simply mean that these are different dwarf galaxies in the
merging process. \citet{leloudas15} showed that hydrogen-poor SLSNe hosts are found in environments
where the gas is strongly ionized, more than in stripped CCSN hosts, and have stronger [\ion{O}{iii}] $\lambda 5007$
emission lines compared to the hosts of other transients types, but comparable to EELGs.

We performed aperture photometry using {\sc SExtractor} \citep{1996A&AS..117..393B},
and we calibrated against $griz$ magnitudes of stars C2, C3, C4, C6 and C7 in
Fig. \ref{sn2002gh_fc_fig} from SDSS. We summarize the host galaxy photometry 
in Table \ref{host_phot_tab}. 

\begin{table}
\caption{DECam $griz$ photometry for the SN\,2002gh host galaxy.}
\label{host_phot_tab}
\begin{tabular}{lcc}
   \hline
   \multicolumn{1}{l}{Filter} &
   \multicolumn{1}{c}{mag} &
   \multicolumn{1}{c}{Abs. mag}  \\
   \hline
$g$ & $24.3$($0.2$)     & $-17.3$ \\
$r$ & $23.5$($0.1$)     & $-18.2$ \\
$i$ & \textgreater$24.1$ & \textgreater$-17.5$ \\
$z$ & $23.3$($0.3$)     & $-18.3$ \\
\hline
\end{tabular}
\begin{tablenotes}
\item Numbers in parentheses correspond to 1\,$\sigma$ statistical uncertainties.
\end{tablenotes}
\end{table}

\subsection{Light curve characterization}
\label{sec:light_curve_chara}

SN\,2002gh was discovered a few weeks before the epoch of maximum bolometric
luminosity (see Section \ref{bolometric:sec}), and our observations begin six days
before bolometric maximum in the rest-frame. We distinguish four phases
in the light curve evolution of SN\,2002gh in Fig. \ref{sn2002gh_lc_fig}. The first phase corresponds
to the light curve evolution close to maximum light, a second phase
of decline post-maximum until 70\,days after maximum, a third phase of faster decline between days 69-105
days, and a fourth phase between 100-205 days, for which we only have observations
in the $V$-band.
 
The optical observations presented here begin shortly before maximum brightness
in the $VRI$ bands, and after $B$-band maximum. The $I$-band observations show
a bumpy structure close to peak, which is caused by a lower signal-to-noise
photometry in this band. To estimate the epoch and the brightness at maximum
light in $VRI$ bands, we used the Gaussian process interpolation. 
An upper limit on the MJD of the $B$-band maximum and the
in the magnitude at this epoch was placed. The uncertainty in the time of maximum light
corresponds to the time difference between the epoch of the maximum from 
the Gaussian process interpolation, and the epoch of the brightest photometric observation
(synthetic photometry for the $R$ band). The uncertainty in the maximum brightness
is the difference between the brightest value in the Gaussian process interpolation and
the brightest photometric observation in the corresponding band. These values are
summarized in Table \ref{tab:peak_mag}.

A first order polynomial was fitted to the post-peak photometric observations
in the $BVI$ filters and to the $R-$band sythetic photometry.  
The goodness-of-fit of these linear models was estimated by computing $\chi^{2}_{\nu}$
and the root mean squared ({\it rms}) of the fit. Inspecting
the fit residuals, we find that a first order polynomial is a suitable model
over these limited ranges of time. The values for the decline rates,
$\chi^{2}_{\nu}$, and the {\it rms} are summarized in Table \ref{tab:linear_decline}.   

During this second phase, spanning from after peak brightness ($t \textgreater 15$\,days)
to 70\,days after bolometric maximum, the SN luminosity declined more rapidly in the bluer bands
with decline rates of 1.82, 1.31, 0.98 and 0.79 mags/100 days in $B$, $V$, $R$, $I$, respectively.
In particular, we notice that the $B$-band declined two times faster than the $I$-band, at the
4\,$\sigma$ level, pointing to a rapid decrease of the UV emission after maximum due to a decrease
of the ejecta temperature, as a consequence of which there is an increase in the strength
of the absorption lines mainly due to \ion{Fe}{ii} and intermediate-mass elements (IME's), as
we discussed above (see Section \ref{sec:spec_analysis}).  

In the third phase, from 69 to 105\,days after maximum brightness, we found that the decline rates
increase to 4.24, 2.95 and 3.32 mags/100 days in $B$, $V$, $I$ bands, respectively. This is an
increase by a factor of 2.3, 2.2 and 4.2, respectively, compared to the decline rates in previous phase
in the same bands. Note the rapid decline in the $B$ band which is larger by $1.3 \pm 0.3$ mag/100 days
(4\,$\sigma$) compared to the $V$-band decline rate. This implies a rapid fading of the near-UV emission.
The $V$ and $I$ decline rates are similar at this phase within 1\,$\sigma$.
We notice that all the decline rates are faster than the expected decline of $\sim 1$ mag/100 days
for a SN powered by the radioactive decay of $^{56}$Co, the daughter product of $^{56}$Ni. The
decline rate in the $V$ band in the fourth phase was measured by fitting a linear model
to the last three photometric observations in this band, finding a value of $1.33 \pm 0.17$
mags/100 days. Although there is a gap of nearly 100 days between the first point and the last two points,
the $V$ band slowed its decline rate to less than half of its previous phase value,
at the 7.5\,$\sigma$ level. This provides evidence of a flattening in the light
curve at this late phase. 

%mu = 41.449

\begin{table}
\caption{Peak magnitude information for SN\,2002gh.}
\label{tab:peak_mag}
\begin{tabular}{@{}lccc}
\hline
Filter & MJD peak & Peak      & Peak abs. \\
       & (days)   & magnitude & magnitude \\
\hline
$B$    & $\leq 52576.8$    & $\leq 19.48$     & $\leq -22.22$ \\
$V$    & $52568.8$($3.4$)  & $19.24$($0.04$) & $-22.40$ \\
$R$    & $52574.8$($2.2$)  & $19.11$($0.05$) & $-22.49$ \\
$I$    & $52569$($26$) & $19.04$($0.09$) & $-22.51$ \\
\hline
\end{tabular}
\begin{tablenotes}
\item Numbers in parentheses correspond to 1\,$\sigma$ statistical uncertainties.
\end{tablenotes}
\end{table}

\begin{table*}
\begin{center}
\caption{Decline rates in the optical bands for SN\,2002gh.}
\label{tab:linear_decline}
\begin{tabular}{@{}lcccccccccccc}
\hline
%Filter & Dec. rate at ph \textless 70\,d & $\chi^{2}_{\nu}$ & $\mathrm{\it rms}$ & $n_{\mathrm{obs}}$ & Dec. rate at 60\,d \textless ph \textless 105\,d & $\chi^{2}_{\nu}$ & $\mathrm{\it rms}$  & $n_{\mathrm{obs}}$ & Dec. rate at ph \textgreater 100\,d & $\chi^{2}_{\nu}$ & $\mathrm{\it rms}$  & $n_{\mathrm{obs}}$ \\
Filter & Dec. rate phase 1 & $\chi^{2}_{\nu}$ & $\mathrm{\it rms}$ & $n_{\mathrm{obs}}$ & Dec. rate phase 2 & $\chi^{2}_{\nu}$ & $\mathrm{\it rms}$  & $n_{\mathrm{obs}}$ & Dec. rate phase 3 & $\chi^{2}_{\nu}$ & $\mathrm{\it rms}$  & $n_{\mathrm{obs}}$ \\
       & (mag/100\,days)                 &                  & (mag)              &                    & (mag/100\,days)                                  &                  & (mag)               &                    & (mag/100\,days)         &   & (mag)               & \\
\hline
$B$    & $1.82$($0.09$) & $0.71$ & $0.030$ & $5$  & $4.24$($0.29$) & --     & --      & $2$ & --             & --     & --      & --  \\
$V$    & $1.31$($0.06$) & $0.04$ & $0.006$ & $5$  & $2.95$($0.17$) & $0.13$ & $0.013$ & $5$ & $1.33$($0.17$) & $0.38$ & $0.064$ & $3$ \\
$R$    & $0.98$($0.98$) & $0.03$ & $0.006$ & $3$  & --             & --     & --      & --  & --             & --     & --      & --  \\
$I$    & $0.79$($0.10$) & $1.99$ & $0.056$ & $5$  & $3.32$($0.29$) & $1.22$ & $0.045$ & $4$ & --             & --     & --      & --  \\
\hline
\end{tabular}
\begin{tablenotes}
\item Numbers in parentheses correspond to 1\,$\sigma$ statistical uncertainties.
\end{tablenotes}
\end{center}
\end{table*}

\subsection{Blackbody fits, bolometric and pseudo-bolometric light curves}
\label{bolometric:sec}

The spectral energy distribution (SED) of hydrogen-poor SLSNe can be well described
by a blackbody model from the time of explosion to a few days or weeks after maximum light \citep[e.g.,][]{nicholl17c}.
After maximum light, the SLSN emission becomes progressively dominated by broad features similar to broad-line SNe Ic
\citep{liu17}, and then enter progressively in the nebular phase \citep{jerkstrand17, nicholl16b, nicholl19}.

\begin{figure}
\centering
\includegraphics[width=85mm]{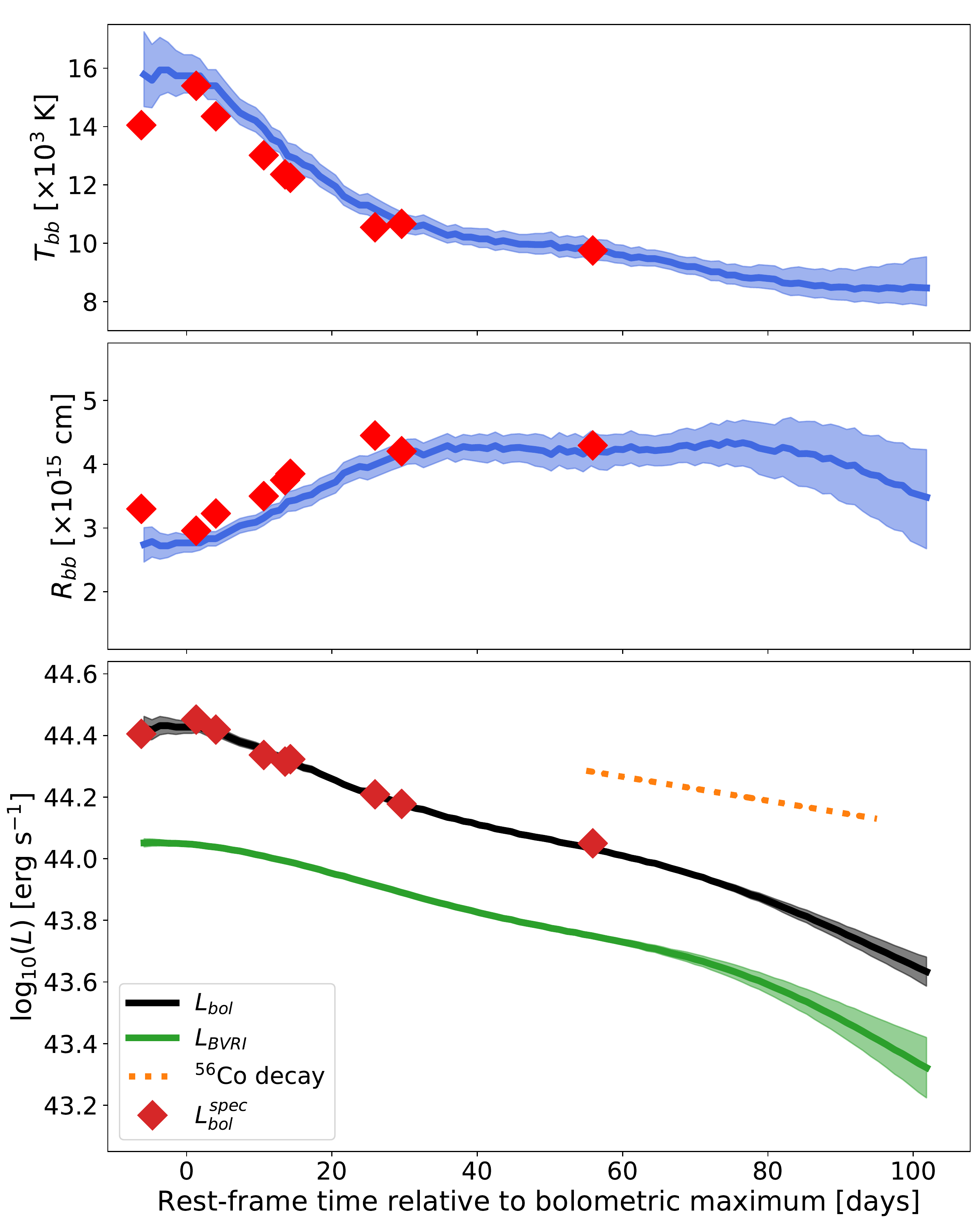}
\caption{Blackbody parameters, bolometric and observed optical luminosity light curves for SN\,2002gh.
In the top and middle panel we present the blackbody temperature $T_{\mathrm{bb}}$ and
radius $R_{\mathrm{bb}}$ obtained from the blackbody fits to the $VRI$ interpolated photometry
of SN\,2002gh (blue line) and to the spectra (red diamonds; see Fig \ref{sn2002gh_spec_fig}).
In the bottom panel, we present the $L_{\mathrm{bol}}$ (black) and $L_{BVRI}$ (green)
light curves and compare them with the decline rate of $^{56}$Co.
We also present the individual points computed from the blackdody fits to the spectra for
$L_{\mathrm{bol}}$ (red diamonds).}
\label{sn2002gh_bb_fit_fig}
\end{figure}

To estimate the radius and the temperature of the SN ejecta we fitted a modified blackbody model to the
interpolated $VRI$ photometry of SN\,2002gh corrected by Galactic extinction. The conversion from magnitudes in
the Vega photometric system to monochromatic fluxes was done using the zero points listed in the Table A2 of
\citet{bessell98}. The modified blackbody function employed to compute the SN bolometric
luminosity, corresponds to a Planck function with a linear flux suppression below the ``cutoff'' wavelength at 3000 \AA\ as shown in Fig.
\ref{sn2002gh_spec_fig} and is described in the Appendix. To complement $V$ and $I$ bands in the blackbody fits, we used $R-$band synthetic
photometry and extrapolated this band by about 50\,days using the Gaussian process model shown in Fig. \ref{sn2002gh_lc_fig}.
We compared fits obtained from $VRI$ and $VI$ photometry, and found that both methods yield similar results.
We noticed that the addition of $R-$band synthetic photometry brings the blackbody parameters closer to
the values estimated from fitting a blackbody to the SN spectra (see Figs. \ref{sn2002gh_spec_fig} and \ref{sn2002gh_bb_fit_fig}).

We did not include $B$-band photometry in our blackbody fits as this band is affected by strong
absorptions from about +14 days after maximum, as can be seen in Fig. \ref{sn2002gh_spec_fig} and discussed
in Sections \ref{sec:light_curve_chara} and \ref{sec:spec_analysis}. To study this effect, we compared
the blackbody parameters obtained by fitting a blackbody model to the $BVRI$ and
to the $VRI$ photometry. At early times (\textless +14 \,days) the $B-$band emission is not
significantly affected by strong spectral features and the derived blackbody parameters obtained
including $B$-band photometry yield an excellent match with the blackbody computed using
$VRI$ photometry and with the SN spectra. After this, the $B$ band becomes progressively affected by the decrease
in the ejecta temperature (Section \ref{sec:spec_analysis}), showing clear differences between
the blackbody temperature and radius computed with and without including the $B$-band. However, in all cases
we find a good agreement in the bolometric luminosity. We decided to use $VRI$ to perfom the
blackbody fits because it yields the most reliable and consistent blackbody parameters at all epochs.

The blackbody fits were performed by $\chi$-square minimization and assuming a luminosity distance of $1,951.7$ Mpc,
which yielded three physical parameters for each epoch: the blackbody temperature ($T_{\mathrm{bb}}$), radius
($R_{bb}$) and the bolometric luminosity. To estimate the uncertainty in these parameters a Monte Carlo simulation
was performed, where we simulate $15,000$ times a new set of photometric points assuming
a Gaussian distribution with the standard deviation equal to the photometric uncertainty, and the
blackbody parameters were re-computed obtaining a distribution for each of them. From the distribution obtained for
$T_{bb}$ and $R_{bb}$, lower and upper $1\sigma$ limits were computed (Fig. \ref{sn2002gh_bb_fit_fig}).
Using the blackbody parameters and their uncertanties, we then computed the bolometric luminosity
($L_{\mathrm{bol}}$) and its uncertainty. The observed optical luminosity ($L_{BVRI}$) was estimated
summing the emission in the $BVRI$ bands (see Fig. \ref{sn2002gh_bb_fit_fig}).

The peak of the bolometric emission was on $\mathrm{MJD} = 52570.8 \pm 2.4$\,days with a luminosity
of $2.6 \pm 0.1 \times 10^{44}$ erg s$^{-1}$ ($log_{10}(L^{\mathrm{peak}}_{\mathrm{bol}}) \simeq 44.4$).
This value places SN\,2002gh among the brightest of SLSNe, with only a handful of objects reaching brighter peak
bolometric luminosities as we discuss below. The peak of the observed luminosity was on $\mathrm{MJD} = 52564.3 \pm 1.5$\,days,
reaching a luminosity of $1.13 \pm 0.03 \times 10^{44}$ erg s$^{-1}$ ($log_{10}(L^{\mathrm{peak}}_{\mathrm{BVRI}}) \simeq 44.0$).
The ratio between the bolometric luminosity and the luminosity observed in the optical bands
is $L^{\mathrm{peak}}_{\mathrm{bol}}/L^{\mathrm{peak}}_{\mathrm{BVRI}} = 2.3 \pm 0.1$.  

We estimate that the total radiated energy from $-6$ to $+102$\,days is $1.2 \times 10^{51}$ erg,
and a total radiated energy observable in the optical bands of $6.1 \times 10^{50}$ erg, over
the same period of time. These values represent a lower limit to the energy emitted by the SN.

The decline rates of the bolometric and pseudo-bolometric light curves over the phase from 15 to 70\,days
and from 70\,days to the end of the light curves ($\simeq 100$\,days) were measured. 
These ranges are equivalent to the second and third phases analysed in the previous section,
however, here we use interpolated light curve values while in Section \ref{sec:light_curve_chara}
we used individual photometric observations.
These luminosity decline rates in magnitudes declined over 100 days are summarized in Table \ref{tab:bol_linear_decline}.
Note that the decline rate during the first phase (15 to 70\,days) is more rapid than the decline rate expected for a
light curve powered by the $^{56}$Co $\rightarrow$  $^{56}$Fe decay, but it could be still consistent with the $^{56}$Co
decay. However, the decline rate measured for the second phase ($t$ \textgreater +70\,days) is 2.4 times faster than
the expected decline rate for a $^{56}$Co powered light curve. This suggests that the decay of $^{56}$Ni
is not the main power source of SN\,2002gh. This will be dicussed in detail in Section \ref{nickel_sec} below. 

While the $L_{\mathrm{bol}}$ computation assumes a blackbody SED, which does
not hold at all epochs, particularly at late phases, $L_{\mathrm{bol}}$ and $L_{BVRI}$ decline rates
are in very good agreement, specially at $t$ \textgreater +70\,days. We also note that
over similar phases the $L_{\mathrm{bol}}$ decline rates are similar to the $V$-band decline
rates presented in Table \ref{tab:linear_decline}.

\begin{table}
%\begin{center}
\caption{Luminosity decline rates of SN\,2002gh.}
\label{tab:bol_linear_decline}
\begin{tabular}{@{}lcc}
\hline
Light curve          & Dec. rate at +15\,d \textless ph \textless +70\,d & Dec. rate at ph \textgreater +70\,d \\
                     & (mag/100\,days)                                   & (mag/100\,days)                    \\
\hline
$L_{\mathrm{bol}}$   & $1.49$($0.02$)                                    & $2.38$($0.11$) \\ 
$L_{BVRI}$           & $1.44$($0.02$)                                    & $2.38$($0.22$) \\
\hline
\end{tabular}
\begin{tablenotes}
\item Numbers in parentheses correspond to 1\,$\sigma$ statistical uncertainties.
\end{tablenotes}
%\end{center}
\end{table}

\subsubsection{Luminosity comparison}
\label{bolometric_comp:sec}

It has been shown that SN\,2002gh is a bright SN belonging to the SLSN class. However, at this point
it is unclear how bright the event is relative to other SLSNe. In this section, the
bolometric light curve of SN\,2002gh is compared with a large sample of SLSNe, placing this object
in the context of the SLSN population as a whole. To this end, a set of bolometric
light curves are computed, using photometry from spectrocopically confirmed hydrogen-poor SLSNe from Pan-STARRS1
\citep[PS1;][]{lunnan18} and the Dark Energy Survey \citep[DES;][]{angus19}. The large PTF SLSN sample
\citep{decia18} is not included in this analysis as their multi-band photometry is not
uniform accross their entire sample, thus making it difficult to construct a homogeneus
set of bolometric light curves. Well observed objects from the literature are not included
in our sample of bolometric light curves either, as individual objects suffer from biases that can be
difficult to account for. Note that the PS1 and DES SLSN samples suffer from spectroscopic selection
biases, inherent to any spectroscopic sample, which are dicussed by \citet{lunnan18} and \citet{angus19},
respectively. Nevertheless, these two SLSN samples represent a well characterized data set, with
well defined photometric systems, providing a uniform and deep multi-band photometry ($griz$), thus perfect for contructing
bolometric light curves. The redshift ranges are $0.3 \lesssim z \lesssim 1.6$
for the PS1 sample and $0.2 \lesssim z \lesssim 2.0$ for the DES sample. 

The rest-frame optical and NIR is the spectral region of choice for fitting a blackbody SED to
SLSNe. For this reason we restrict ourselfs to construct bolometric light curves for objects at
$z \lesssim 1.2$. This choice is justified in the Appendix.

In the left-panel of Fig. \ref{bol_and_obs_comp_fig}, the bolometric light curves of 28 SLSNe-I,
11 from PS1 and 17 from DES, are presented and compared with the bolometric light curve of SN\,2002gh. As can
be observed SN\,2002gh reaches one the largest maximum luminosities of this sample, with only two SNe from PS1 reaching
similar values. Note that PS1 SLSNe are concentrated in the bright region of the distribution compared with DES objects,
having a maximum bolometric luminosity range of $log_{\mathrm{10}}(L_{\mathrm{bol}}) = 43.5-44.4$.
The DES SLSNe span a wider luminosity range with $log_{\mathrm{10}}(L_{\mathrm{bol}}) = 43.25-44.3$.
This is likely explained in part by the fact that the DES survey limiting magnitude, in terms of photometry 
and spectroscopy, reaches nearly 1 mag fainter than the PS1 survey. As can be noticed in the figure there is
a large diversity in the shape of the light curves, with some objects having early bumps and
others showing bumps after maximum light. We also find a large diversity in the rise times with objects having
rise times of about 30 days and others showing broader light curves, with PS1-14bj being an extreme object with
a rise time of more than 100 days. No apparent relation between the peak luminosity and the light
curve shape is found.

\begin{figure*}
\centering
\includegraphics[width=180mm]{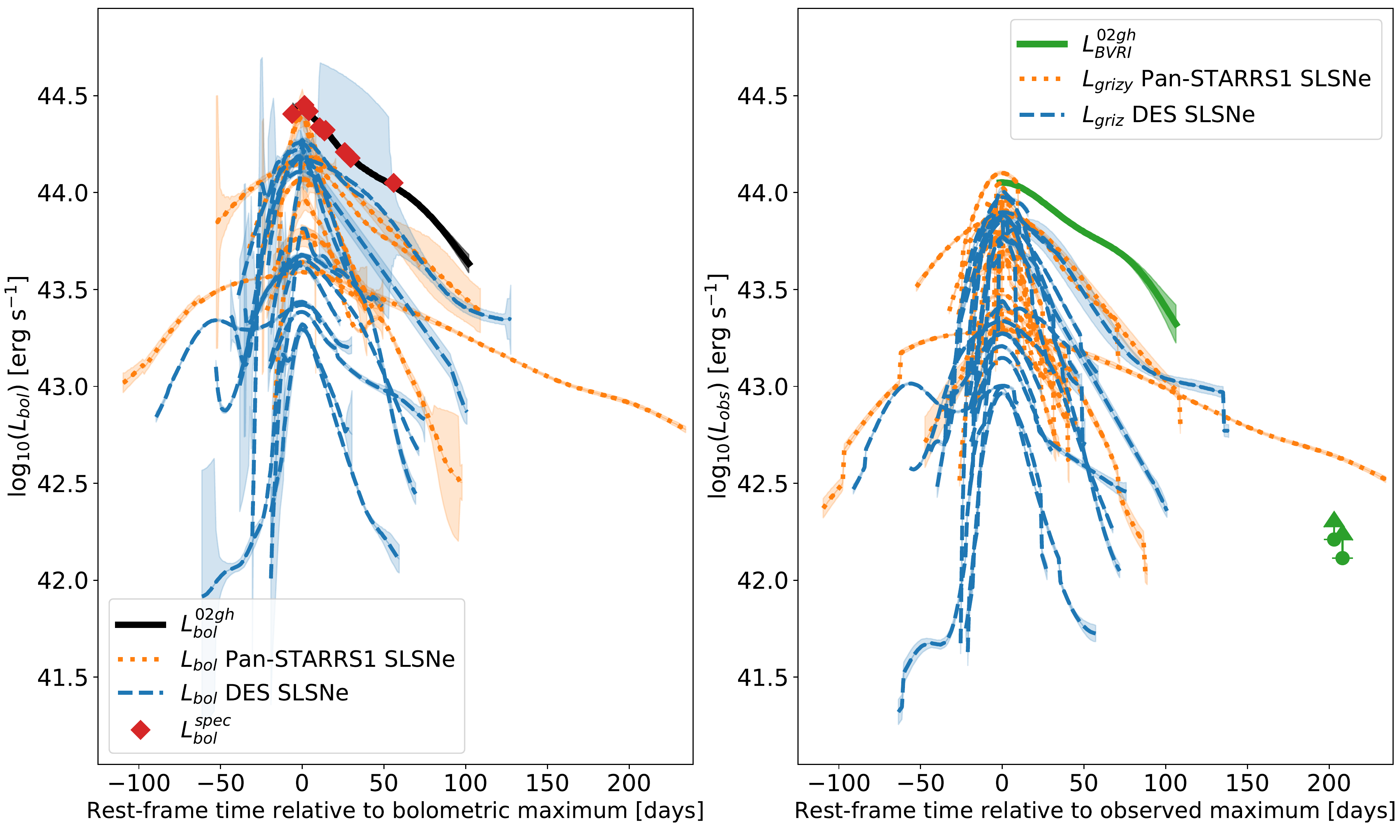}
\caption{Bolometric and observed luminosity light curve
comparison between SN\,2002gh and the spectroscopically
confirmed SLSNe from Pan-STARRS1 \citep{lunnan18} and
DES \citep{angus19}. In the left panel, we present the bolometric
light curves and in the right panel we present the emitted luminosity
detected in the optical bands. In the Appendix we describe our methodology
to compute bolometric and observed luminosity light curves in detail.
}
\label{bol_and_obs_comp_fig}
\end{figure*}

In the right-panel of Fig. \ref{bol_and_obs_comp_fig} the luminosity light curves for
PS1 and DES SLSNe constructed by adding the emission detected in the optical bands are presented.
Here the sample is not restricted to objects with $z \lesssim 1.2$, thus all PS1 and DES objects
are included.
SN\,2002gh is again among the brightest SLSNe, only PS1-13or ($z=1.53$) reaches a brighter maximum
luminosity. Here again PS1 SLSNe are concentrated in a brighter region compared with DES objects,
which are distributed over a wider luminosity range.

\begin{figure}
\centering
\includegraphics[width=85mm]{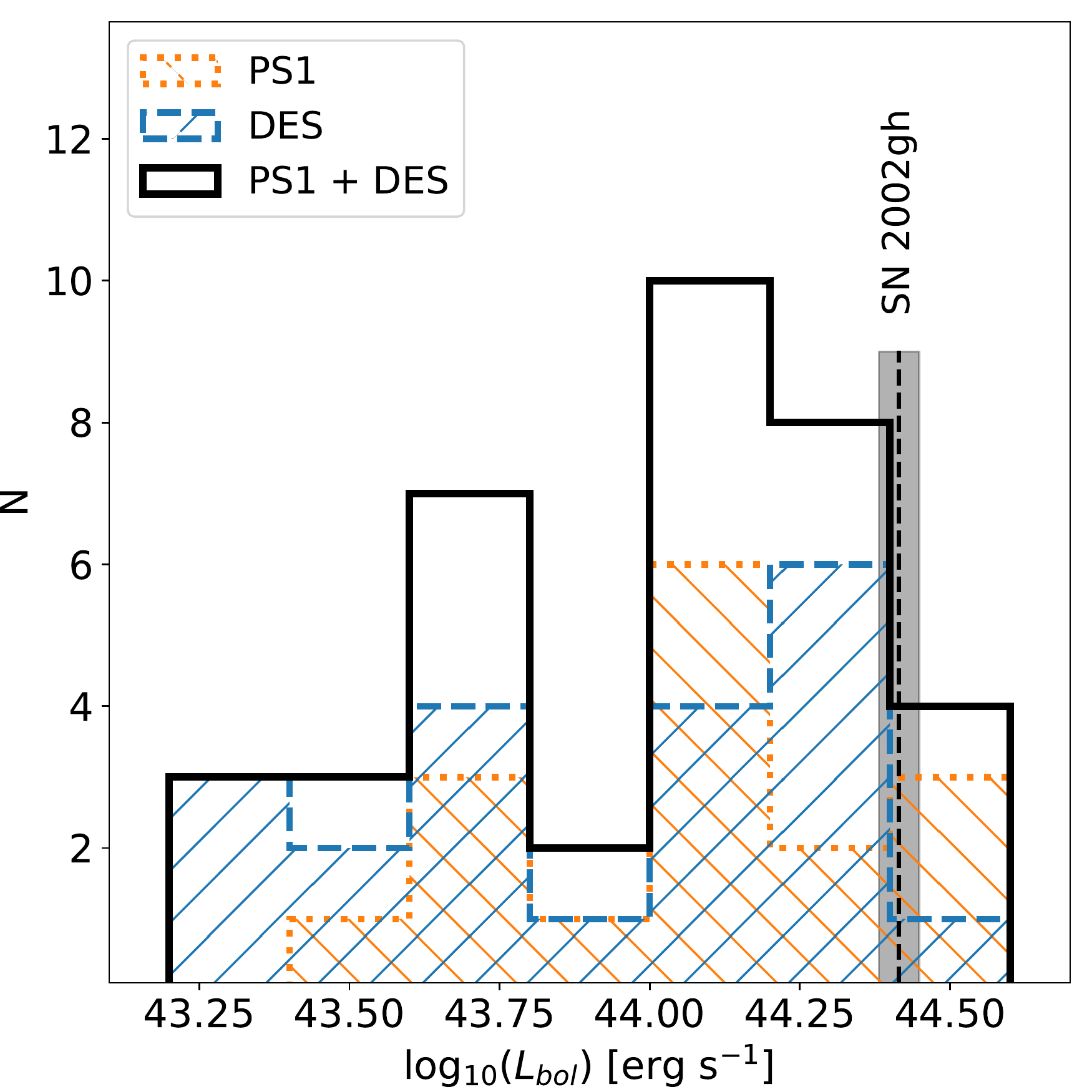}
\caption{Observed maximum bolometric luminosity distribution for spectroscopically
confirmed SLSNe from Pan-STARRS1 \citep{lunnan18} and
DES \citep{angus19}. This distribution includes peak bolometric luminosity for
objects at $z \textgreater 1.2$, which are estimated using a heuristic method described
in the Appendix. We mark with a black dashed line the maximum bolometric luminosity of SN\,2002gh,
confirming that it is among the most luminous SNe ever observed.
}
\label{Lbol_distributions_fig}
\end{figure}

In Fig. \ref{Lbol_distributions_fig} the maximum bolometric luminosity distribution is presented
for spectroscopically confirmed hydrogen-poor SLSNe from PS1 and DES. This distribution includes
all possible PS1 and DES objects, 37 SLSNe in total. For objects at $z \textgreater 1.2$, we estimate
the maximum bolometric luminosity using the heuristic $L_{\mathrm{bol}}/L_{\mathrm{obs}}$ scaling relation
described in the Apendix. Only PS1-10awh ($z=0.909$), PS1-11bam ($z=1.565$), PS1-13or ($z=1.53$),
and DES15E2mlf ($z=1.861$) have a similar or brighter maximum luminosity than SN\,2002gh. This figure
confirms that SN\,2002gh lies at the extreme of the distribution, only comparable
to the most luminous objects in a sample of 37 SLSNe, which includes the most distant and luminous
objects detected to date. 

\subsection{Power source}
\label{power_source_sec}

In this Section the magnetar model together with radioactive decay of $^{56}$Ni
are explored as possible sources to power the extreme luminosity of SN\,2002gh.
We also discuss potential signatures of interaction between the SN ejecta
and the CSM.

\subsubsection{Magnetar model}
\label{magnetar_sec}

The power injection from the spin down of a newborn magnetar is a theoretical model
able to explain the extreme luminosities of hydrogen-poor SLSNe. Under the assumption that SN\,2002gh is
powered by the spin down of a magnetar, here we use the magnetar model implemented by \citet{nicholl17c}
to fit the photometry of SN\,2002gh using two methodologies: 1) the Modular Open Source Fitter for Transients
\citep[{\sc MOSFiT};][]{guillochon18} that uses the SN photometry as input, and 2) our {\sc python} implementation
of the model to fit the bolometric light curve directly.

Here we briefly summarize the main analytic expresions of the magnetar model of \citet{nicholl17c}.
We refer the reader to this article and the references therein for a detailed description of the
model. The magnetar energy input ($F_{\mathrm{mag}}$) is:

\begin{equation}
 F_{\mathrm{mag}}(t) = \frac{E_{\mathrm{mag}}}{t_{\mathrm{mag}}} \frac{1}{(1+t/t_{\mathrm{mag}})^{2}},
\end{equation}

\noindent where the magnetar rotational energy ($E_{\mathrm{mag}}$) is

\begin{equation}
 E_{\mathrm{mag}} =  2.6 \times 10^{52} \left( \frac{M_{\mathrm{NS}}}{1.4~M_{\sun}} \right)^{3/2} \left( \frac{P_{\mathrm{spin}}}{1~\mathrm{ms}} \right)^{-2} \mathrm{erg,} 
\end{equation}

\noindent the magnetar spin down timescale ($t_{\mathrm{mag}}$) is given by

\begin{equation}
 t_{\mathrm{mag}} = 1.3 \times 10^{5} \left( \frac{M_{\mathrm{NS}}}{1.4~M_{\sun}} \right)^{3/2} \left( \frac{P_{\mathrm{spin}}}{1~\mathrm{ms}} \right)^{2} \left( \frac{B}{10^{14}~\mathrm{G}} \right)^{-2} \mathrm{s.} 
\end{equation}

\noindent In these expressions $M_{\mathrm{NS}}$, $P_{\mathrm{spin}}$, and $B$ correspond to the mass, spin period, and magnetic field of the neutron star, respectively.

The output SN luminosity ($L_{\mathrm{SN}}$) is calculated using the traditional analytic solution of \citet{arnett82}, where in this case
the SN energy source is $F_{\mathrm{mag}}(t)$ and the analytic expression takes the following form:

\begin{equation}
 L_{\mathrm{SN}}(t) = e^{-(t/t_{\mathrm{diff}})^2} (1 - e^{-A t^{-2}}) \int\limits_0^t 2~F_{\mathrm{mag}}(t') \frac{t'}{t_{\mathrm{diff}}} e^{(t'/t_{\mathrm{diff}})^2} \frac{dt'}{t_{\mathrm{diff}}}, 
\end{equation}

\noindent where the diffusion timescale is given by

\begin{equation}
 t_{\mathrm{diff}} = \left( \frac{2 \kappa M_{\mathrm{ej}}}{\beta c v_{\mathrm{ej}}} \right)^{1/2},
\end{equation}

\noindent and includes a leakage parameter \citep{wang15, nicholl17c}, not included in the original \citet{arnett82} analytic expression, of the form:

\begin{equation}
 A =  \frac{3 \kappa_{\gamma} M_{\mathrm{ej}}}{4 \pi v^2_{\mathrm{ej}}}.
\end{equation}

\noindent In these expressions the parameter $\beta$ has the typical value of 13.7 \citep{arnett82, inserra13}, 
$c$ is the speed of light, $M_{\mathrm{ej}}$ is the ejecta mass, $v_{\mathrm{ej}}$ is the
characteristic ejecta velocity, $\kappa$ is the optical opacity, and $\kappa_{\gamma}$ is the
opacity to high-energy photons. The leakage term, $(1 - e^{-A t^{-2}})$, controls the fraction of high-energy
photons thermalized by the SN ejecta \citep{wang15, nicholl17c}. A large value of $\kappa_{\gamma}$
implies a significant thermalization of the magnetar input energy at all times, while
a small $\kappa_{\gamma}$ value implies that a large fraction of the magnetar energy escapes
in the form of high energy radiation at late times \citep{wang15, vurm21}, when the ejecta density drops.

\begin{figure}
\centering
\includegraphics[width=85mm]{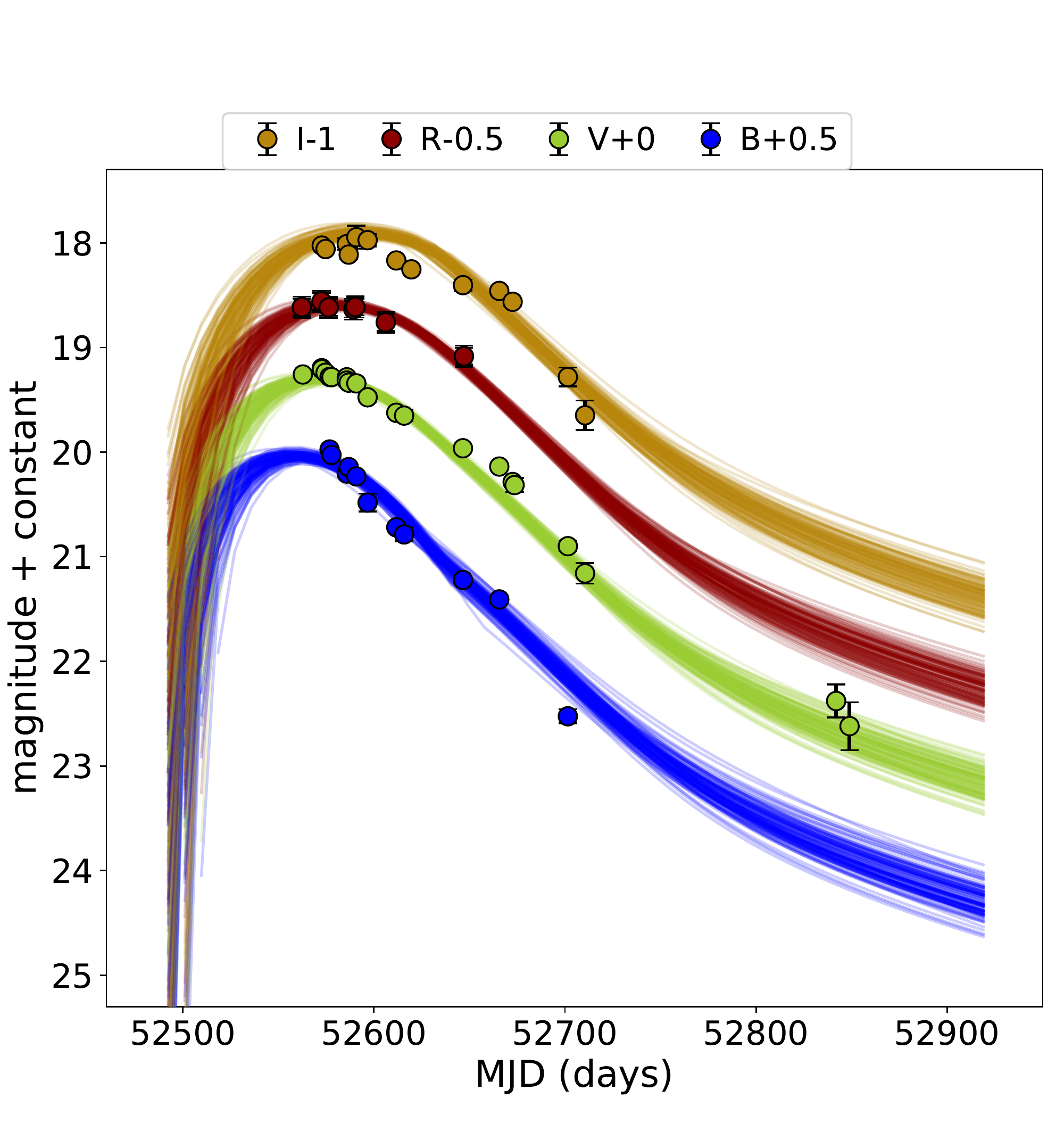}
\caption{Magnetar model fit to SN\,2002gh photometry using the \citet{nicholl17c} implementation in
{\sc MOSFiT} \citep{guillochon18}.}
\label{mosfit_fig}
\end{figure}

In Fig. \ref{mosfit_fig} we present the {\sc MOSFiT} magnetar model fit result for SN\,2002gh,
where we fixed $z=0.365$ and the Galactic reddening to $E(B-V) = 0.0593$\,mag in the fit. We
do not allow {\sc MOSFiT} to fit for host galaxy extinction or the gas column density parameters.
Note that this fit describes reasonably well the light curve evolution, including 
the late $V$-band photometry, with the most significant discrepancies in the $B$ and $I$ bands.
We used additional information obtained from the analysis of SN\,2002gh spectra presented
in Section \ref{sec:spec_analysis} to set restrictions on $v_{\mathrm{ej}}$.
In Section \ref{sec:spec_analysis}, we measured an ejecta expansion velocity between
$9000$ and $11,000$ km\,s$^{-1}$ over a rest-frame time of more that 60\,days. Over this time interval
we probed several layers of the SN ejecta, quantifying changes in the ejecta temperature and velocity.
Therefore, it is expected that our estimate for the SN expansion velocity should be representative
of the ejecta velocity. For this reason, a Gaussian prior on $v_{\mathrm{ej}}$ is employed,
having a mean velocity of $10,000$ km\,s$^{-1}$ and a standard deviation of $1000$ km\,s$^{-1}$.
Additionally, the following restriction $7,000 \textless v_{ej} \textless 16,000$ km\,s$^{-1}$ was imposed.
Without these restrictions a value of $v_{ej} \simeq 5800$\,km\,s$^{-1}$ would be obtained.
This $v_{\mathrm{ej}}$ value represents 60\% of the expansion velocity estimated from the minima of the
spectral features in Section \ref{sec:spec_analysis}.
The best {\sc MOSFiT} values the for SN\,2002gh are summarized in Table \ref{tab:magnetar_params}.
We highlight the large ejecta mass of $M_{ej} \simeq 25\, M_{\sun}$ obtained by {\sc MOSFiT}, and note
that similar $M_{\mathrm{ej}}$ values are obtained regardless of the restrictions on $v_{\mathrm{ej}}$.

\begin{figure*}
\centering
\includegraphics[width=180mm]{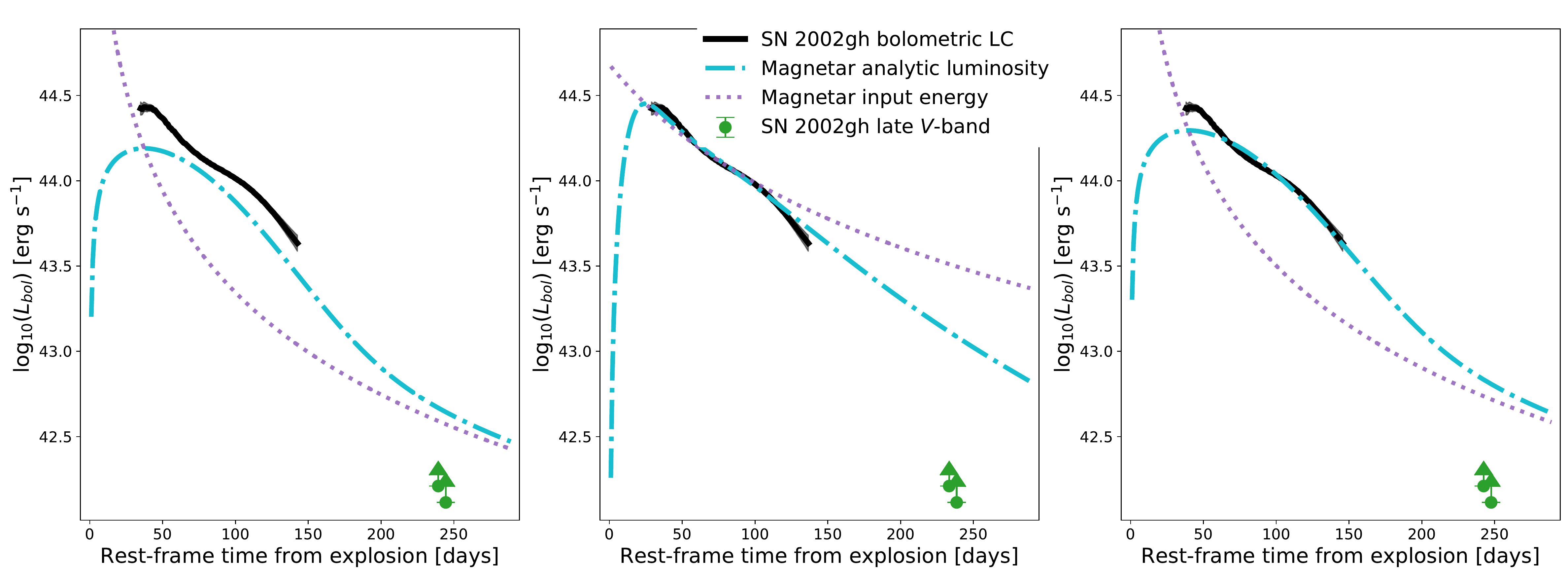}
\caption{Left panel: comparison between the {\sc MOSFiT} analytic bolometric light curve (cyan dash-dotted line) and the observed
bolometric light curve of SN\,2002gh (black continuous). Note that {\sc MOSFiT} fits the magnetar model to the photometry directly,
and not to the observed bolometric light curve of SN\,2002gh shown in this panel. In all the panels, the purple dotted line represents
the magnetar input energy to the SN ejecta, and the green circles are the late time $V$-band photometry of SN\,2002gh which provide a
lower limit for the SN bolometric luminosity at this phase. Middle panel: best magnetar fit to the bolometric
light curve of SN\,2002gh using our grid exploration method and including leakage. Right panel: best magnetar fit to the bolometric
light curve of SN\,2002gh without energy leakage.
}
\label{analytic_boloLC_fig}
\end{figure*}

Fig. \ref{analytic_boloLC_fig} shows the analytic luminosity for the magnetar model using the {\sc MOSFiT}
parameters obtained for SN\,2002gh, and compared with the bolometric light curve computed for SN\,2002gh in the
previous Section. There is an overall agreement between the two, but with the {\sc MOSFiT} analytic light curve shifted
by 0.10 to 0.15 dex to fainter luminosities than our estimated bolometric light curve. This shift is the result
of differences in the methodologies employed to fit the modified blackbody model to the photometry and in the fitting
procedure of the magnetar model itself. {\sc MOSFiT} uses a Markov chain Monte Carlo (MCMC) fitter, to fit 
simultaneously for all the parameters in their model, including the photospheric velocity ($v_{\mathrm{phot}}$) and the
final plateau temperature ($T_{f}$) which determine the blackbody temperature and radius evolution. In our case the
evolution of the temperature and radius are not parameterized but fitted at each epoch independently. Another difference
is the inclusion of the $B$ band photometry in the fit, which in our case does not yield
significant differences in the bolometric emission, but which under a different fitting methodology and a different
implementation of the modified blackbody emission may be a source of systematic discrepancies. 
Additionally, the {\sc MOSFiT} light curve fit includes the late time $V-$band photometry at $200$\,days, while
the bolometric light curve ends at about $100$\,days after the bolometric maximum. The two methodologies use a slighly different
version of the Galactic reddening law to account for the Galactic extiction, which may introduce a small but systematic
difference in the same direction observed. The {\sc MOSFiT} analytic luminosity light curve peaks at about five days
before the estimated epoch of the bolometric maximum,
having a bolometric maximum luminosity of $1.5 \times 10^{44}$ erg\,s$^{-1}$ ($log_{\mathrm{10}}(L_{\mathrm{bol}}) = 44.17$).
This value represents 57.7\% of the maximum luminosity estimated for SN\,2002gh. Despite the constraints
imposed on $v_{\mathrm{ej}}$, {\sc MOSFiT} still tends to converge to a low ejecta velocity.

Motivated by these small but apparent discrepancies, we decided to explore a wide parameter space
and fit directly the bolometric light curve of SN\,2002gh using consistently the \citet{nicholl17c}
magnetar model implementation. To simplify the model and to reduce the number of parameters to fit,
we decided to fix the velocity to $v_{\mathrm{ej}} = 10,000$ km\,s$^{-1}$, and the neutron star mass
to $M_{\mathrm{NS}} = 1.7$\,$M_{\sun}$ for a direct comparison with the {\sc MOSFiT} fit. We also discuss
the case of $M_{\mathrm{NS}} = 1.4$\,$M_{\sun}$ below. These values represent reasonable simplifiying
assumptions. Our strategy consists in assessing a wide range of parameters that reasonably reproduce
the bolometric light curve of SN\,2002gh. To this we used a grid over a large range
in $M_{\mathrm{ej}}$, $B$, $P_{\mathrm{spin}}$, and $\kappa$ and performed a least-squares
minimization for $t_{\mathrm{exp}}$, and $\kappa_{\gamma}$ parameters to fit the bolometric
light curve of SN\,2002gh. The ranges explored for $M_{\mathrm{ej}}$, $B$, and $P_{\mathrm{spin}}$ are
$0.63 \leq M_{\mathrm{ej}} \textless 50.0~M_{\sun}$ with a logarithmic step
of $log_{\mathrm{10}}(\Delta M_{ej}) = 0.1$, $0.1 \leq B \textless 12.5 \times 10^{14}~\mathrm{G}$ 
with a logarithmic step of $log_{\mathrm{10}}(\Delta B) = 0.1$, $0.60 \leq P_{\mathrm{spin}} \textless 7.25~\mathrm{ms}$
with a linear step of $\Delta P_{\mathrm{spin}} = 0.20$\,ms, and $ 0.1  \leq \kappa \leq 0.2$\,cm$^{2}$\,gr$^{-1}$ with 
a linear step of  $\Delta \kappa = 0.025$\,cm$^{2}$\,gr$^{-1}$
respectively. Our motivation here is to explore the magnetar parameter space and find the sets of values
able to reproduce the bolometric light curve of SN\,2002gh, not only the set that yields the best fit.

The $\chi^{2}_{\nu}$ computed for each set of parameters was used to find good fits
satisfying the energy balance requirement, that is
$E_{K} \textless E_{\mathrm{mag}} - E_{\mathrm{rad}} + E_{\nu}$ \citep[see][]{nicholl17c}.
Where $E_{K}$ ($= \frac{1}{2}M_{\mathrm{ej}} v^{2}_{\mathrm{ej}}$) is the kinetic energy
of the SN ejecta, $E_{\mathrm{mag}}$ is the magnetar rotational energy, $E_{\mathrm{rad}}$
is the energy radiated and $E_{\nu} \approx 10^{51}$ erg is the canonical energy of a
core-collapse explosion. Then, we selected the fits satisfying the energy balance and that
also have $\chi^{2}_{\nu} \textless 20$. This $\chi^{2}_{\nu}$ value was selected after
visual inspection of the fits, providing a good threshold for reasonable fits. 

Our exploration found two clearly distinct groups in the parameter space clustered around lower $\chi^{2}_{\nu}$ values,
that reproduce the bolometric light curve of SN\,2002gh. These two groups can be mainly distinguished by their ejecta mass and the need of a
leakage parameter. The best magnetar parameters for these solutions are summarized in Table \ref{tab:magnetar_params}.
We found that the best fits correspond to an ejecta mass in the range of 0.6 to 3.2\,$M_{\sun}$, $\kappa_{\gamma}$
between 0.16 and 0.53\,cm$^{2}$\,gr$^{-1}$, $\kappa$ between 0.1 and 0.2\,cm$^{2}$\,gr$^{-1}$,
$P_{\mathrm{spin}}$ in the range 2.8-3.4\,ms, and $B = 0.50 \times 10^{14}$\,G. These fits have a $\chi^{2}_{\nu}$ between
2.02 and 2.18, and $\kappa$ varies proportional to $\kappa_{\gamma}$ and inversely proportional to the ejecta mass. The best
fit is presented in the middle-panel of Fig. \ref{analytic_boloLC_fig}. The maximum luminosity of this model
is $2.8 \times 10^{44}$\,erg\,s$^{-1}$, consistent with our estimate, the rise time to maximum luminosity
is 24.9\,days, this corresponds to about 10\,days before the estimated epoch of maximum bolometric luminosity in the rest-frame.
Starting at about 100\,days after the explosion, we see that a noticeable fraction of the magnetar input energy
leaks from the ejecta without being thermalized. This epoch corresponds to $60-70$\,days after our estimated
epoch of bolometric maximum (MJD=52570.8), and corresponds to the phase of rapid decline of the bolometric light curve. 

The second set of solutions corresponds to a local minima and is characterized by a huge ejecta mass between about 30 and 40 $M_{\sun}$, 
having $\kappa_{\gamma} \textgreater 1.0$\,cm$^{2}$\,gr$^{-1}$. This set of solutions has problems
to reproduce the maximum of the bolometric light curve as is also the case of the {\sc MOSFiT} fit
(see Fig. \ref{analytic_boloLC_fig}), having as a consequence larger $\chi^{2}_{\nu}$ values.
Note that the huge ejecta mass together with $\kappa_{\gamma} \textgreater 0.1$\,cm$^{2}$\,gr$^{-1}$, implies that the energy
leakage is negligible and these solutions are equivalent to the case of no leakage. In these fits, the spin period is between 0.6
and 1.0\,ms, the magnetic field is between 1.6 and 2.0$\times 10^{14}$\,G, $\kappa$ is between 0.175 and 0.2\,cm$^{2}$\,gr$^{-1}$, and the
best solutions for this case have $\chi^{2}_{\nu} \simeq 10$. The rise time from the SN explosion to maximum luminosity
is about 40\,days, which corresponds to about six days before the estimated epoch of the bolometric maximum. This
model has a maximum luminosity of about $2.0 \times 10^{44}$\,erg\,s$^{-1}$. The parameters for this case
are in good agreement with the {\sc MOSFiT} values, within the uncertainties (see Table \ref{tab:magnetar_params}). The
agreement can be considered even better if the systematic differences between the two methodologies employed
are considered. Hence, we consider the {\sc MOSFiT} solution equivalent to our best bolometric fit without leakage.

The mass function of neutron stars show evidence of a bimodal distribution, with a low-mass component centred 
at $1.4 M_{\sun}$, and a high-mass component centred at $1.8 M_{\sun}$ \citep{antoniadis16}. To investigate
the effect of the neutron star mass in our results we fitted the bolometric light curve of SN\,2002gh
fixing $M_{\mathrm{NS}} = 1.4 M_{\sun}$. We found that both solutions remain in this case, and as in
the case of $M_{\mathrm{NS}} = 1.8 M_{\sun}$ the second solution is found when $\kappa$ is between
0.175 and 0.2\,cm$^{2}$\,gr$^{-1}$. In Table \ref{tab:magnetar_params} we summarize the parameters
for the best fits in each case. The values change slightly compared with the $M_{\mathrm{NS}} = 1.7 M_{\sun}$
case, but are still consistent with the previous results.   
 
\begin{table*}
\caption{Parameters of the magnetar model fits of SN\,2002gh.}
\label{tab:magnetar_params}
\renewcommand{\arraystretch}{1.4}
\begin{tabular}{@{}lcccccccc}
\hline
Magnetar parameters          & $M_{\mathrm{ej}}$     & $v_{\mathrm{ej}}$             & $P_{\mathrm{spin}}$ & $B$                  & $\kappa$               & $\kappa_{\gamma}$     & $M_{\mathrm{NS}}$   & $\chi^{2}_{\nu}$ \\
                             & ($M_{\sun}$)          & ($\times 10^{3}$ km s$^{-1}$) & (m\,s)              & ($\times 10^{14}$ G) & (cm$^{2}$ gr$^{-1}$)   & (cm$^{2}$ gr$^{-1}$)  & ($M_{\sun}$)        &             \\
\hline
{\sc MOSFiT}                 & $25.1^{+5.1}_{-3.7}$  & $7.8^{+0.6}_{-0.6}$           & $1.2^{+0.2}_{-0.2}$ & $1.9^{+3.8}_{-1.1}$  & $0.17^{+0.02}_{-0.03}$ & $6.0^{+125.8}_{-5.7}$ & $1.7^{+0.2}_{-0.3}$ & --          \\ 
Bolometric fit no leakage    & $31.6$                & $10.0^{\dagger}$              & $1.0$               & $1.6$                & $0.200$                & $\infty$              & $1.7^{\dagger}$     & 10.1       \\
                             & $31.6$                & $10.0^{\dagger}$              & $0.6$               & $1.2$                & $0.200$                & $\infty$              & $1.4^{\dagger}$     & 10.9       \\
Bolometric fit with leakage  & $2.0$                 & $10.0^{\dagger}$              & $3.2$               & $0.5$                & $0.100$                & $0.22$                & $1.7^{\dagger}$     & 2.0        \\
                             & $1.0$                 & $10.0^{\dagger}$              & $3.0$               & $0.4$                & $0.100$                & $0.38$                & $1.4^{\dagger}$     & 1.6        \\
\hline
\end{tabular}
\begin{tablenotes}
\item $^{\dagger}$Fixed values.
\end{tablenotes}
\end{table*}

%Mns =1.4
%23.57 10000.0 0.1 0.38007 1.0   3.0 0.3981 164.3 1.65981 0.994235 4.21160412782 1.92881144689 2.28855768092
%48.56 10000.0 0.2 1.0 31.6 0.6 1.259 1088.0 10.9939 31.417826 105.290103195 3.58463847307 71.2876387224

%$log_{\mathrm{10}}(M_{\mathrm{ej}})$, $log_{\mathrm{10}}(B)$ and $P_{\mathrm{spin}}$ in steps of 0.10, 0.10, and 0.20, respectively.

\subsubsection{Radioactive decay of $^{56}$Ni}
\label{nickel_sec}

Here we explore the radioactive decay of $^{56}$Ni as the possible power source for the light curve
of SN\,2002gh. The input energy from the $^{56}$Ni radioactive decay \citep[see][]{nadyozhin94} is: 

\begin{equation}
 F_{\mathrm{Ni}}(t) = \left( 6.45 \times 10^{43} e^{-t/\tau_{Ni}} + 1.45 \times 10^{43} e^{-t/\tau_{Co}} \right) \frac{M_{\mathrm{Ni}}}{M_{\sun}}  \mathrm{erg}\,\mathrm{s,}^{-1}
\label{Ni_decay_eq}
\end{equation}

\noindent where $\tau_{Ni} = 8.8$\,days and $\tau_{Co} = 111.3$\,days are the e-folding decay times of $^{56}$Ni and $^{56}$Co, respectively.

\begin{figure}
\centering
\includegraphics[width=85mm]{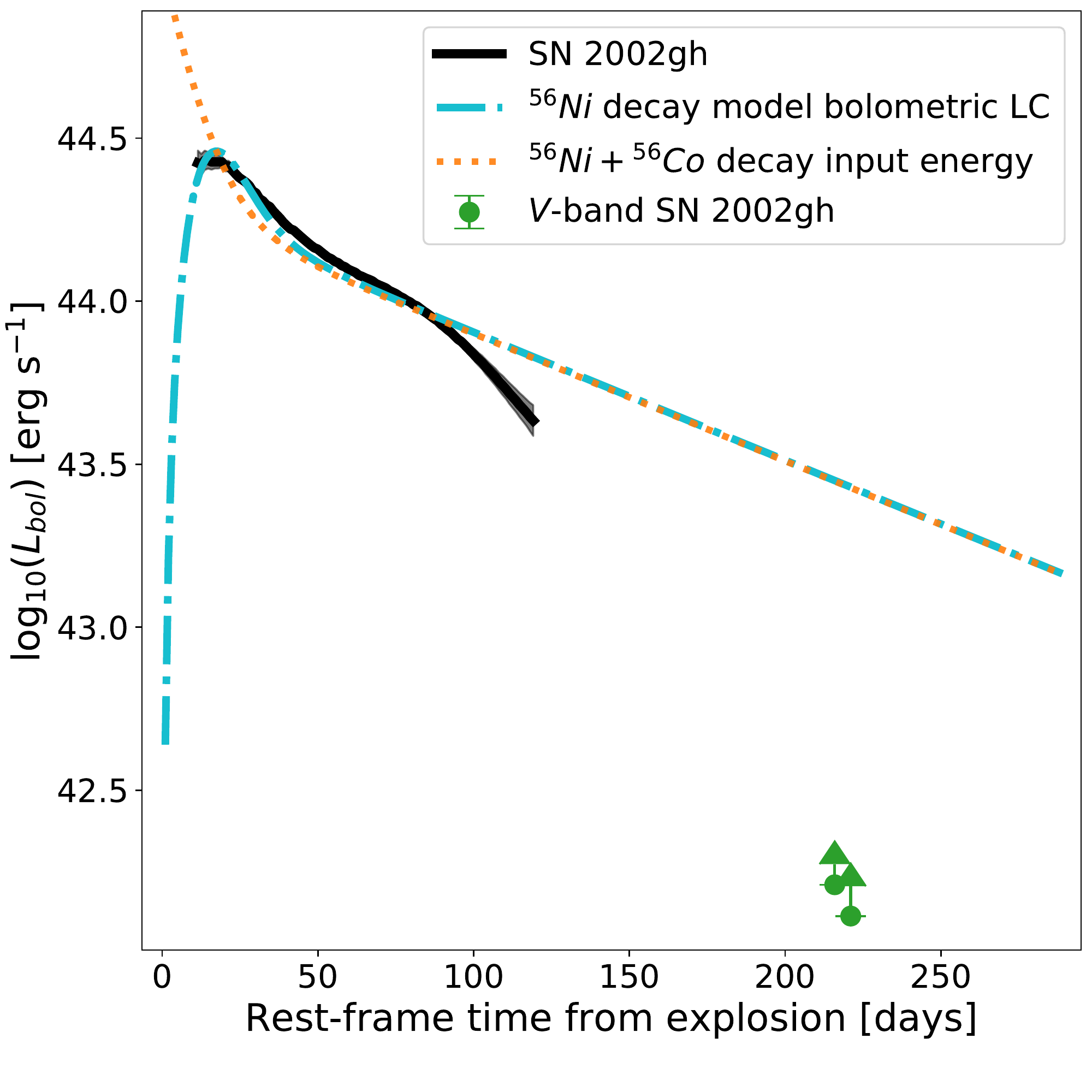}
\caption{ Fit of the radioactive decay of $^{56}$Ni (Eq. \ref{Ni_decay_eq})
to the bolometric light curve of SN\,2002gh, using the \citet{arnett82}
analytic approximation.
The parameters of the best fit model are $\tau_{\mathrm{rise}} = 17.3$\,days,
$M_{\mathrm{ej}} = 0.9 M_{\sun}$, $M_{\mathrm{Ni}} = 13.6 M_{\sun}$,
$v_{\mathrm{ej}}= 10,000$ km\,s$^{-1}$ and $\kappa = 0.2$ cm$^{2}$\,gr$^{-1}$.
This model illustrates that the light curve of SN\,2002gh follows
an evolution {\it ``consistent''} with the $^{56}$Ni decay until about 70-80 days post
maximum. However, the best fit obtained with the $^{56}$Ni decay model is non-physical as
$M_{\mathrm{Ni}} \textgreater M_{\mathrm{ej}}$, thus we can rule out the $^{56}$Ni decay
as the main power source for SN\,2002gh.
}
\label{Ni_decay_boloLC_fig}
\end{figure}

Using the \citet{arnett82} analytic expresion we fitted the bolometric light curve of SN\,2002gh,
finding that the best fit requires a total ejecta mass of $M_{\mathrm{ej}} = 0.9 M_{\sun}$, a
$^{56}$Ni mass of $M_{\mathrm{Ni}} = 13.6 M_{\sun}$, and has a rise time of 17.3 days, as shown in
Fig. \ref{Ni_decay_boloLC_fig}. This is an non-physical solution as the total ejecta mass is smaller than the total $^{56}$Ni mass.
Increasing the rise time also increases the $M_{\mathrm{Ni}}$ necessary to fit the light curve, but we always find that
$M_{\mathrm{Ni}} \textgreater M_{\mathrm{ej}}$. Since the total ejecta mass cannot be smaller than the $^{56}$Ni mass required
to power the SN luminosity, we conclude that SN\,2002gh maximum luminosity cannot be powered by the $^{56}$Ni radioactive decay.
Although small amounts of $^{56}$Ni (of about $\sim 0.1 M_{\sun}$) must be synthesized during the explosion, this energy source does
not dominate the light curve evolution of SN\,2002gh.

\subsubsection{Circumstellar medium interaction}
\label{sec:csm_inter}

An alternative scenario to explain the observed luminosities of SLSNe is the interaction between
the SN ejecta an a massive CSM \citep[see][ and references therein]{moriya18}. The smoking gun evidence
for ejecta-CSM interaction in SLSNe-I is the presence of broad hydrogen or helium emission lines
\citep[see e.g.,][]{yan15,yan17,fiore21,pursiainen22}. Alternatively, a bumpy structure in the SN light
curve could be a signature of CSM interaction as well \citep[see e.g.,][]{yan17,fiore21}.  

In the case of SN\,2002gh we do not find clear signatures of ejecta-CSM interaction. In particular,
the spectra of SN\,2002gh do not show any signature of interaction (see Section \ref{sec:spec_analysis}).
Inspecting carefully the early $I-$band bumpy structure (Section \ref{sec:light_curve_chara}), we find
that the residuals between the photometric observations and the light curve interpolation are correlated
with the photometric uncertainty. No correlation is found between the $I-$band residuals and the residuals
in the other bands as would be expected for the case of CSM interaction. The $B$ and $V$ light curves show
a smooth evolution, without signatures of bumps.
 
The potential signatures of ejecta-CSM interaction are the break in the $BVI$ light curves at
about 70 days after maximum light and the light curve flattening at late times. The light curve break
does not exhibit an increase in the SN luminosity as is usually observed in bumps produced by the ejecta-CSM
interaction \citep[see][]{yan17,fiore21}. In SN\,2002gh the break consists in an increase of the decline rates in all bands
(see Section \ref{sec:light_curve_chara}), potentially explained by the leakage of high-energy radiation
\citep[see Section \ref{magnetar_sec};][]{wang15,vurm21}. Unfortunately, we do not have a spectrum
of SN\,2002gh at the time of the break or at the time of the flattening to detect spectral features
associated with a potential ejecta-CSM late time interaction.

\section{Discussion and conclusions}
\label{sec:discussion}

We have presented and analysed excellent quality observations of SN\,2002gh obtained as part of the CATS project. SN\,2002gh
is at a redshift of $z=0.365 \pm 0.001$, it was the second SLSN-I discovered after SN\,1999as, and its true nature remained
unnoticed for two decades. The multi-band observations presented here for SN\,2002gh cover 210\,days in the rest-frame
of the SN evolution, with a well sampled spectral sequence comprising 60\,days. These observations make of SN\,2002gh
a perfect object for a detailed study, such as the one presented here.

In Section \ref{sec:spec_analysis} we study the spectral evolution of SN\,2002gh, identifying several
ions at an expansion velocity decreasing from $11,000$ in the early phase to
$9000$ km\,s$^{-1}$ in the later phases. We compare the spectra of SN\,2002gh with well observed hydrogen-poor
SLSNe and stripped envelope core-collapse SNe from the literature. The spectral evolution of SN\,2002gh is typical
of an object of this class displaying \ion{O}{ii} lines before and close to maximum light, following a cooling phase,
where the \ion{Fe}{ii} lines become prominent features. After about +26\,days the SN starts to resemble to Type Ic SN\,1994I
or to Type Ic-BL SN\,2002ap at an early phase. SN\,2002gh shows significant spectroscopic similarity to iPTF13ajg,
which is also a bright object of this SN class \citep{vreeswijk14}.

In Section \ref{sec:host_galaxy} we identified the potential SN\,2002gh's host galaxy in deep DECam images. The potential
host is a faint dwarf galaxy, presumably having low metallicity as is common for a SLSN-I hosts \citep{lunnan14, leloudas15, perley16, schulze18},
located 0\farcs13 south-east from the SN location corresponding to a projected distance of 0.67 kpc, at the SN redshift. The host galaxy candidate
shows irregular morphology and multiple emission peaks. The multi-peak morphology could be tracing regions of strong
star-formation within the irregular host or could be interpreted as a system of dwarf galaxies in a merging process. Recently,
\citet{vanggaard20} found that hydrogen-poor SLSN host galaxies are often part of interacting systems, with $\sim 50$\% having at
least one major companion within 5\,kpc. They interpreted this as SLSNe-I being the result of a recent burst of star formation,
possibly triggered by galaxy interaction. Future confirmation that the SN\,2002gh's host galaxy is part of a system of dwarf galaxies in a merging
process, would provide further support to the \citet{vanggaard20} hypothesis. Further deep spectroscopy of the host galaxy candidate is required
for a solid association with the SN and to understand the nature of the multi-peak morphology.  

In Sections \ref{sec:light_curve_chara} and \ref{bolometric:sec} we characterized the light curves of SN\,2002gh and
computed its bolometric light curve assuming that the SN emission can be well described by a modified blackbody.
We distinguished four phases in the light curves of SN\,2002gh. The first phase associated with the maximum light and the immediate decline from
maximum, a second phase from about two weeks to $\simeq +70$\,days post maximum, which is characterized by a bolometric decline
rate roughly consistent with the decline rate expected for a SN powered by $^{56}$Co decay.
A third phase, from about $+70$\,days to $\simeq 100$\,days, characterized by a faster decline rate in all bands. The
last phase comprises the last three photometric observations in the $V$-band, with a gap of nearly a hundred days between the first
and the last two points. We find that the average decline rate value over this phase is reduced compared to the decline rate of the previous phase,
indicating a possible flattening of the light curve at about 200\,days after maximum. 

In Section \ref{bolometric:sec} we find that the maximum bolometric luminosity of SN\,2002gh is $2.6 \pm 0.1 \times 10^{44}$ erg\,s$^{-1}$ 
($log_{10}(L^{\mathrm{peak}}_{\mathrm{bol}}) \simeq 44.4$). Our study presents a large set of bolometric light
curves for 28 SLSNe-I, 11 from Pan-STARRS1 \citep{lunnan18} and 17 from DES \citep{angus19}, showing the large diversity in shapes, including the
presence of pre- and post-maximum bumps, and a large range in maximum luminosities from $log_{10}(L_{\mathrm{bol}}) \simeq 43.2$ to
$log_{10}(L_{\mathrm{bol}}) < 44.6$. A comparison of the bolometric light curve of SN\,2002gh with these objects shows that SN\,2002gh
is the brightest in this sample. Comparing the estimated maximum bolometric luminosity for a total sample of 37 SLSNe-I,
16 from Pan-STARRS1 and 21 from DES, we show that SN\,2002gh is among the most luminous SNe ever observed.

In Section \ref{power_source_sec} we study the magnetar model and the radioactive decay of $^{56}$Ni as power sources of SN\,2002gh.
We rule out the radioactive decay of $^{56}$Ni as the main power source for the light curve of SN\,2002gh, finding that a minimum 
of $\simeq 13 M_{\sun}$ of $^{56}$Ni are needed to power the peak luminosity of SN\,2002gh, an unrealistically large value. Moreover,
the best analytic fitting to the bolometric light curve always requires a total ejecta mass below the $^{56}$Ni mass, which is a 
non-physical solution, thus ruling out the radioactive decay of $^{56}$Ni as the main power source for SN\,2002gh. Although 
some amount of $^{56}$Ni must be synthesized during the SN explosion. 

We found that the spin down of a magnetar can explain the huge maximum luminosity observed in SN\,2002gh. Fitting an
analytic magnetar model to the observed bolometric light curve of SN\,2002gh we found two alternative solutions. The first
solution, requires significant leakage of the magnetar input energy (see Fig. \ref{analytic_boloLC_fig}), of which the
rapid decline observed in the third phase of the SN light curves could be a signature. We found that the best fit to the
bolometric light curve is for $M_{\mathrm{ej}}$ between 0.6 and 3.2 $M_{\sun}$, $P_{\mathrm{spin}} = 3.2$\,ms, and $B=5 \times 10^{13}$\,G. 
The alternative magnetar model does not require power leakage, and it is characterized by a huge ejecta mass of
25 to 40 $M_{\sun}$, a fast spin period of $P_{\mathrm{spin}}$ between 0.6 and 1.2\,ms, and $B$ between 1.6 and 2.0$\times 10^{14}$\,G.     

SN\,2002gh lacks signatures of hydrogen or helium mixed in the SN ejecta, therefore it must have lost all its hydrogen
and a significant part of its helium envelope through winds, but also through binary interaction, since half or more of all massive
stars are found in binary systems. Using equations 13 and 14 of \citet{woosley19} for evolution models of massive helium stars
in binary systems of solar metallicity, and assuming a final mass ($M_{\mathrm{fin}}$) before the explosion between 2.3 and
4.9 $M_{\sun}$ and of about 30 $M_{\sun}$ for the first and second case of the magnetar model, respectively.
We estimate a zero-age main-sequence mass between 14 and 25 $M_{\sun}$ for the first case and of about 135 $M_{\sun}$ for
the second case. These estimates for the main-sequence masses are particularly uncertain at very high mass, due to the
uncertainty in the mass loss prescription \citep[see][]{woosley19}, and the sensitivity of the mass loss to metallicity. Although
we provide just a crude mass estimate for the second case, this estimate would place the progenitor star of SN\,2002gh among the
most massive stars observed to explode as a SN.

We do not find evidence that the ejecta-CSM interaction is the power source of the maximum luminosity of SN\,2002gh,
although late time interaction cannot be discounted (see Section \ref{sec:csm_inter}). The spectral features of SN\,2002gh are
typical of a non-interacting SN, resembling non-interacting stripped envelope core-collapse SNe from about +24\,days to the last spectrum
at about +54\,days (see Fig. \ref{spec_comp_coolphotos_fig}) relative to the bolometric maximum. However, we cannot rule out
completely the case of ejecta-CSM interaction in SN\,2002gh, but we can place some constraints on the potential CSM composition,
such as the CSM must be hydrogen and helium poor. We also note that the magnetar model can explain reasonably well the
observations of SN\,2002gh without the need to invoke the additional power contribution from the ejecta-CSM interaction.

\section*{Acknowledgments}

We thank the anonymous referee for their thorough comments that helped to improve this manuscript. 
We also thank to Matt Nicholl for his prompt and helpful response about {\sc MOSFiT} inquiries, and to
Michael Wood-Vasey for his very helpful comments on the NEAT observations of SN\,2002gh. 
This research draws upon DECam data as distributed by the NOIRLab Astro Data Archive. NOIRLab is
managed by the Association of Universities for Research in Astronomy (AURA) under a cooperative agreement
with the National Science Foundation. MDS is supported by grants from the VILLUM FONDEN (grant number 28021) and
the Independent Research Fund Denmark (IRFD; 8021-00170B).

This project used data obtained with the Dark Energy Camera (DECam), which was constructed by the Dark Energy Survey
(DES) collaboration. Funding for the DES Projects has been provided by the US Department of Energy, the US National
Science Foundation, the Ministry of Science and Education of Spain, the Science and Technology Facilities Council of
the United Kingdom, the Higher Education Funding Council for England, the National Center for Supercomputing Applications
at the University of Illinois at Urbana-Champaign, the Kavli Institute for Cosmological Physics at the University of Chicago,
Center for Cosmology and Astro-Particle Physics at the Ohio State University, the Mitchell Institute for Fundamental Physics and Astronomy at Texas
A\&M University, Financiadora de Estudos e Projetos, Funda\c{c}\~{a}o Carlos Chagas Filho de Amparo \`a Pesquisa do Estado do Rio de Janeiro,
Conselho Nacional de Desenvolvimento Científico e Tecnol\'ogico and the Minist\'erio da Ci\^{e}ncia, Tecnologia e Inova\c{c}\~{a}o,
the Deutsche Forschungsgemeinschaft and the Collaborating Institutions in the Dark Energy Survey.

The Collaborating Institutions are Argonne National Laboratory, the University of California at Santa Cruz,
the University of Cambridge, Centro de Investigaciones En\'ergeticas, Medioambientales y Tecnol\'ogicas–Madrid,
the University of Chicago, University College London, the DES-Brazil Consortium, the University of Edinburgh, the
Eidgen\"ossische Technische Hochschule (ETH) Z\"urich, Fermi National Accelerator Laboratory, the University of
Illinois at Urbana-Champaign, the Institut de Ci\`encies de l'Espai (IEEC/CSIC), the Institut de F\'isica d'Altes Energies,
Lawrence Berkeley National Laboratory, the Ludwig-Maximilians Universit\"at M\"unchen and the associated Excellence
Cluster Universe, the University of Michigan, NSF’s NOIRLab, the University of Nottingham, the Ohio State University,
the OzDES Membership Consortium, the University of Pennsylvania, the University of Portsmouth, SLAC National Accelerator Laboratory,
Stanford University, the University of Sussex, and Texas A\&M University.

Based on observations at Cerro Tololo Inter-American Observatory, a program of NOIRLab (NOIRLab Proposal ID \#2014B-0404; PIs: David Schlegel and Arjun Dey
and NOIRLab Proposal ID \#2019A-0305; PI: Alex Drlica-Wagner), which is managed by the Association of
Universities for Research in Astronomy (AURA) under a cooperative agreement with the National Science Foundation.

\section*{Data Availability}

The photometric data of SN\,20202gh is available in the article and the spectra are available on request to the
corresponding author and also available from the WISeREP archive\footnote{\url{http://wiserep.weizmann.ac.il/}} \citep{yaron12}.

\bibliographystyle{mnras}
\bibliography{references} % if your bibtex file is called example.bib

\appendix

\section{Bolometric light curves}

\begin{figure}
\centering
\includegraphics[width=85mm]{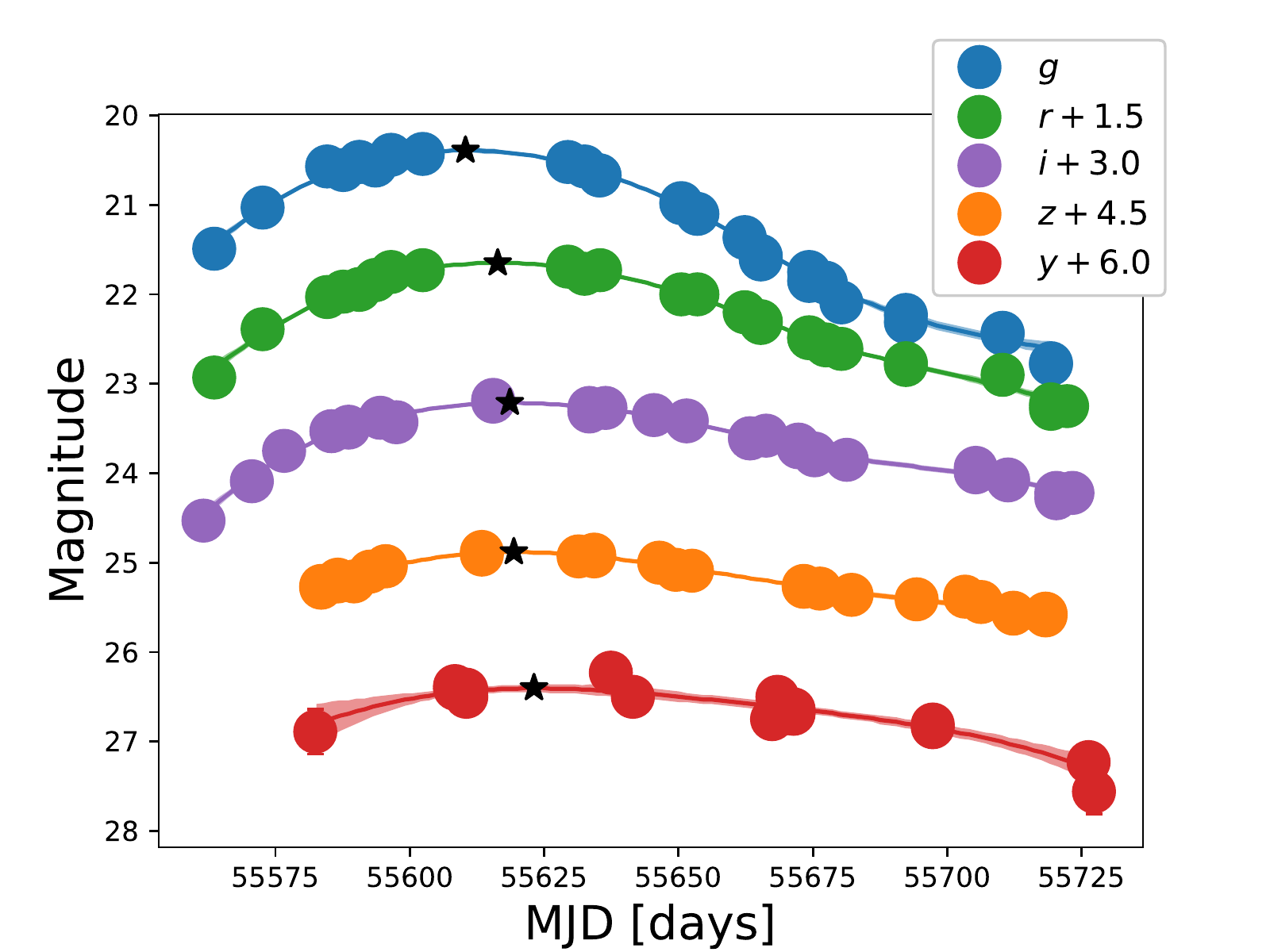}
\caption{Gaussian process light curve interpolation
for PS1-11ap. Continous lines correspond to the light
curve interpolation and shaded regions to 1\,$\sigma$
uncertainties. Black stars mark the peak brightness
for each filter.}
\label{PS1_11ap_interpol_fig}
\end{figure}

\begin{figure}
\centering
\includegraphics[width=85mm]{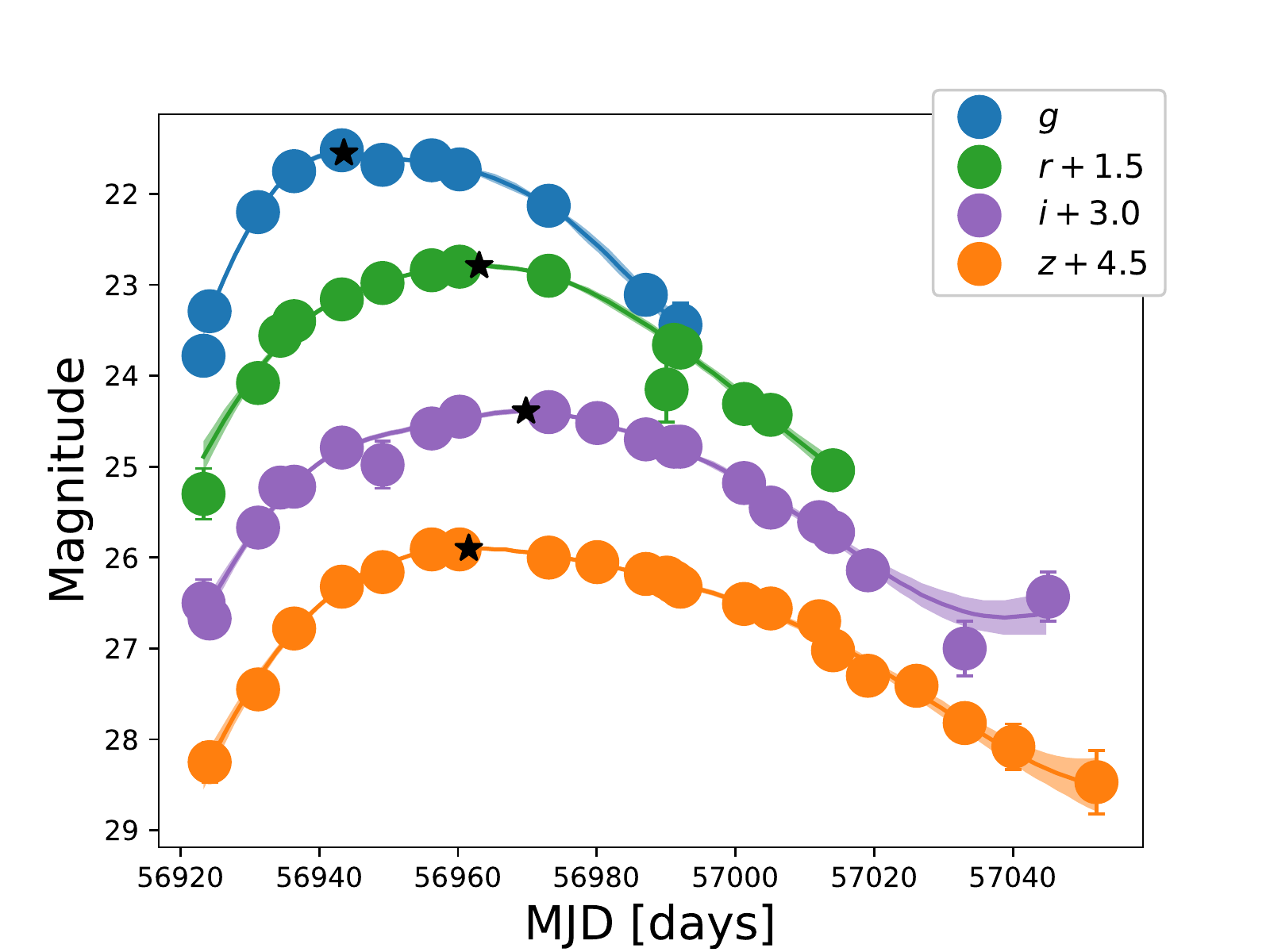}
\caption{Gaussian process light curve interpolation
for DES14X2byo. Continous lines correspond to the light
curve interpolation and shaded regions to 1\,$\sigma$
uncertainties. Black stars mark the peak brightness
for each filter.
}
\label{DES_14X2byo_interpol_fig}
\end{figure}

We constructed bolometric light curves from Pan-STARRS1 \citep[PS1;][]{lunnan18} and DES \citep{angus19} SLSNe following
a procedure identical to the one used for SN\,2002gh. First we interpolated their light curves using the {\sc python}
Gaussian process module implemented in {\sc scikit-learn} \citep{scikit-learn}, using the RBF and Matern kernels. The PS1
dataset is composed of observations in $grizy$ reaching a depth slightly fainter than $\sim 23.3$ mag in $griz$ and $\sim 22$ in $y-$band.
Only a handful of objects have $y-$band photometry, usually more noisy than on the other filters. DES present a homogeneous set of photometry
in $griz$ passbands to a depth of $\sim 24$ mag, obtained with the Dark Energy Camera \citep{flaugher15} which is mounted on the 4\,meter
Blanco telescope at Cerro Tololo in Chile. In figures \ref{PS1_11ap_interpol_fig} and \ref{DES_14X2byo_interpol_fig} we present two examples of
the light curves interpolation, for PS1-11ap (Fig. \ref{PS1_11ap_interpol_fig}) and for DES14X2byo (Fig. \ref{DES_14X2byo_interpol_fig}).
One of the advantages of using Gaussian process interpolation is its flexibility, providing an excellent fit to the SLSNe light curves,
as can be observed in these examples. At the same time, this approach provides a robust estimate for the uncertainty in the interpolation.
Using this approach we obtained an homogeneous set of 37 multi-band interpolated light curves with their respective
uncertainties, the latter point is key for the determination of the uncertainties in blackbody parameters and in our bolometric light curves.

\begin{figure}
\centering
\includegraphics[width=85mm]{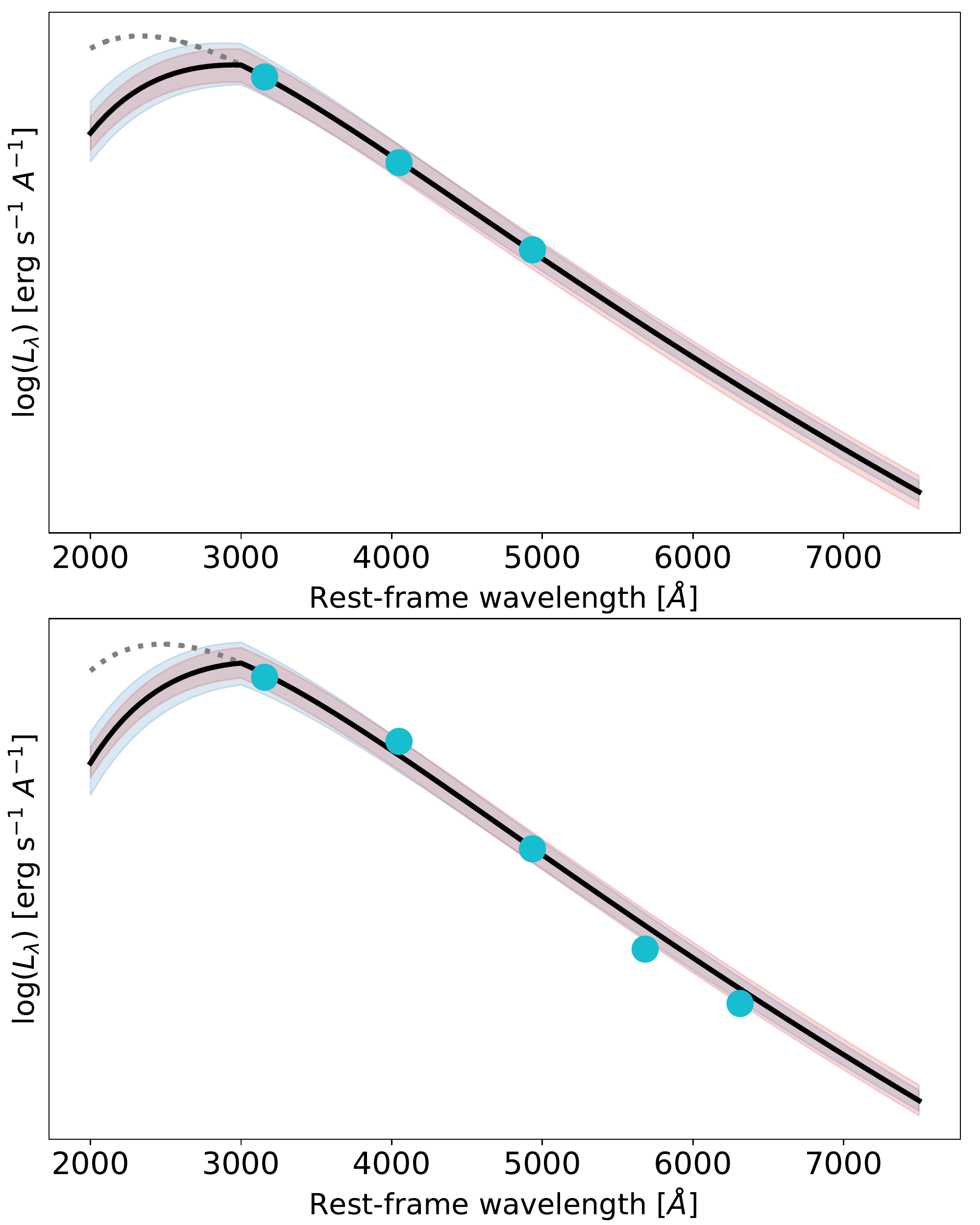}
\caption{Blackbody fits to the interpolated photometry
of PS1-11ap at -22.5\,days relative to bolometric maximum
(top) and at bolometric maximum (bottom). The solid line
is the modified blackbody SED with a linear flux suppression
below the ``cutoff'' wavelength at 3000 \AA. The dotted line
shows the full blackbody for comparison. Shaded blue regions
correspond to 1\,sigma uncertainty in $T_{bb}$ ($R_{bb}$ fixed)
and shaded red regions correspond to 1\,sigma uncertainty
in $R_{bb}$ ($T_{bb}$ fixed).    
}
\label{bb_fit_PS1_11ap_fig}
\end{figure}

\begin{figure}
\centering
\includegraphics[width=85mm]{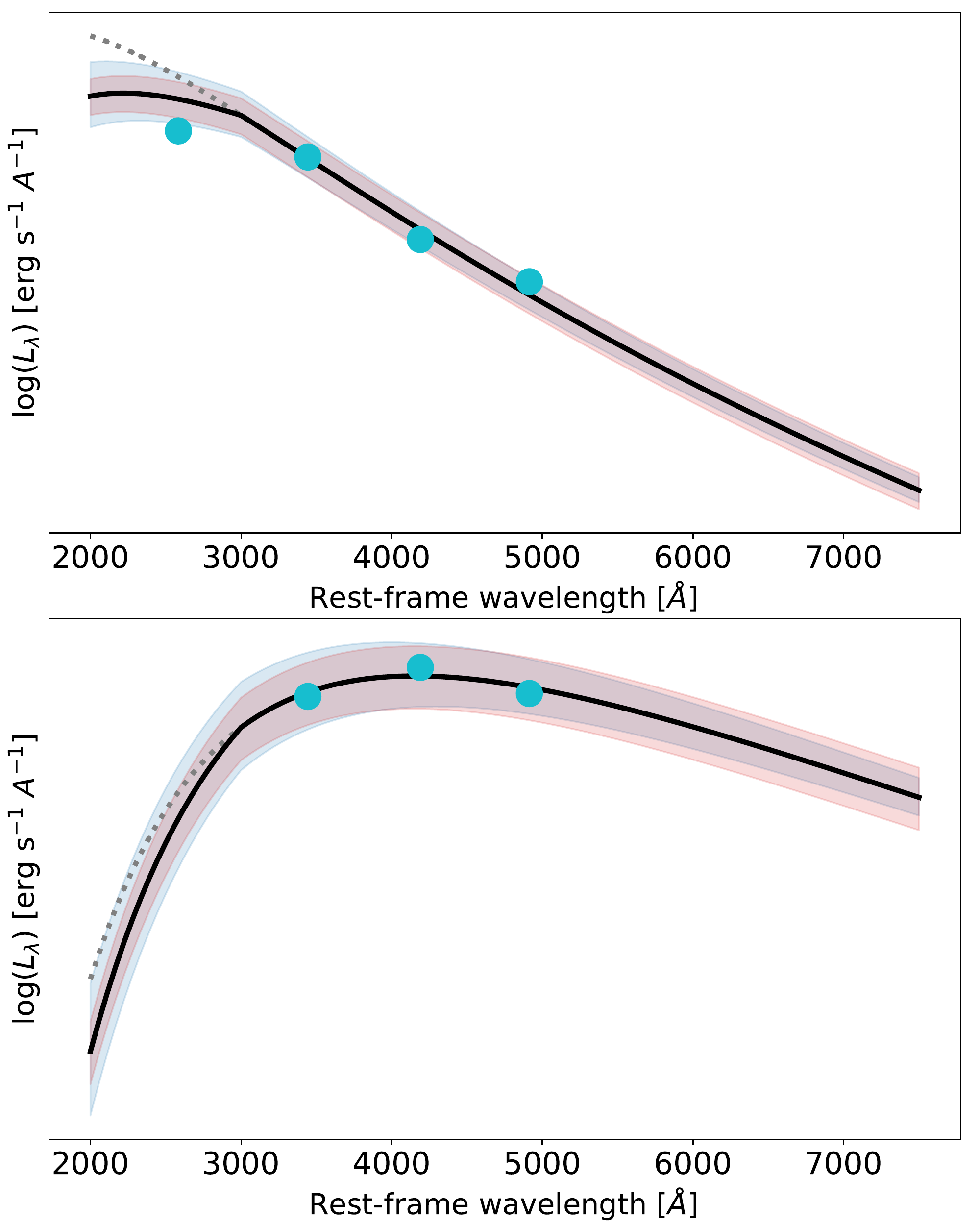}
\caption{Blackbody fits to the interpolated photometry
of DESX214byo at bolometric maximum (top) and at +22.5
relative to bolometric maximum (bottom). The solid line is the
modified blackbody SED with a linear flux suppression
below the ``cutoff'' wavelength at 3000 \AA. The dotted line
shows the full blackbody for comparison. Shaded blue regions
correspond to 1\,sigma uncertainty in $T_{bb}$ and shaded red
regions correspond to 1\,sigma uncertainty in $R_{bb}$.
}
\label{bb_fit_DES_14X2byo_fig}
\end{figure}

To model the SLSN emission we fitted the interpolated light curves corrected for Galactic extinction using a modified blackbody SED.
To correct the light curves by Galactic extinction we used $E(B-V)$ values from \citet{schlafly11} and the \citet{cardelli89}
reddening law with $R_{V} = 3.1$, and no correction for host galaxy extinction was attempted. We converted magnitudes in the AB system
to monochromatic fluxes following the \citet{fukugita96} definition \citep[see also, e.g.,][]{tonry12}. The effective wavelength and full-width at half
maximum were computed for each filter using the DES and PS1 response functions, and we find good agreement with the values reported in \citet{tonry12}
for PS1. The modified blackbody SED corresponds to the Planck function multiplied by a linear flux suppression below the ``cutoff'' wavelength
at 3000 \AA\ as shown in figures \ref{bb_fit_PS1_11ap_fig} and \ref{bb_fit_DES_14X2byo_fig}, where the linear flux suppression corresponds to
a value between zero and one, with a value of zero at zero wavelength and one at 3000 \AA. A similar modified SED has been repeatedly used in the literature
\citep[see e.g.,][]{quimby13,nicholl17c}. Despite the fact that the linear flux suppression below 3000 \AA\ brings the blackbody emission
in better agreement with photometric and spectroscopic observations of SLSNe \citep[see][]{nicholl17c}, the UV-region is dominated by broad
absorption lines that are far from being a perfect modified blackbody. The deviation from this modified blackbody SED is expected to be larger
in the UV region than in the optical, since at optical wavelengths broad absorption lines are shallow and the continuum dominates the SN emission until
a few days or weeks after maximum. Thus, the rest-frame optical and NIR emission provides a better region to fit the underlying blackbody emission.
As our bolometric light curves are based on a robust determination of the blackbody parameters ($T_{bb}$ and $R_{bb}$), we place the
followinng requirements: 1) we only include SLSNe with $z$\textless 1.2 in order to have at least one band, usually the $z$-band,
located at a rest-frame wavelength greater than 4000 \AA, where the {\it `blackbody continuum'} is well defined in SLSNe and 2) the
rest-frame effective wavelength of a band must be greater than 2900 \AA\ to be considered in the fit. To fit a modified blackbody SED a
minimum of two photometric bands is sufficient, but we generally required photometry in at least three bands. Only in cases where the blackbody
fits with two photometric points shows a consistent evolution with blackbody fits obtained using more bands at adjacent epochs, we keep them.
We noticed that this is usually the case for fits perfomed at a phase of before or close to maximum light, when the SLSN emission
is very similar to a blackbody, and two photometric points with wavelength greater than $\sim 3500$ \AA\ can constrain very well the blackbody
parameters. This is no longer true at late times, a few weeks after maximum, when the SLSN emission deviates from a blackbody and two photometric
points are not able to constrain the underlying  pseudo-continuum. In these cases the fits are disregarded, sometimes even using three photometric
observations we obtain fits that are not good and coherent with SED fits at previous epochs, in these cases the blackbody fits are disregarded as well.

To estimate the uncertainty in the fitted blackbody parameters a similar procedure as for SN\,2002gh is followed. A new set of $15,000$ photometric observations
is simulated assuming a Gaussian distribution with the standard deviation equal to the photometric uncertainty estimated using
Gaussian process interpolation, and the blackbody parameters are re-computed obtaining a distribution for each of them. From the distribution obtained
the uncertainties for $T_{bb}$ and $R_{bb}$ are estimated, computing lower and upper 1\,$\sigma$ limits. Then, the bolometric emission using
a blackbody SED with flux suppression below the ``cutoff'' wavelength is computed, and its corresponding 1\,$\sigma$ uncertainty estimated.

In addition to the bolometric light curves, observed luminosity ($L_{\mathrm{obs}}$) light curves were constructed. To construct these light curves we summed
the detected emission in the optical bands using our light curve interpolation method. When constructing these observed luminosity light curves care was put
to take into account the effect of redshift in the effective wavelength and the width of the PS1 and DES filters. In tables \ref{tab:bol_PS1} and \ref{tab:bol_DES}
we summarize the maxima of the observed luminosity light curves. 

\begin{figure}
\centering
\includegraphics[width=85mm]{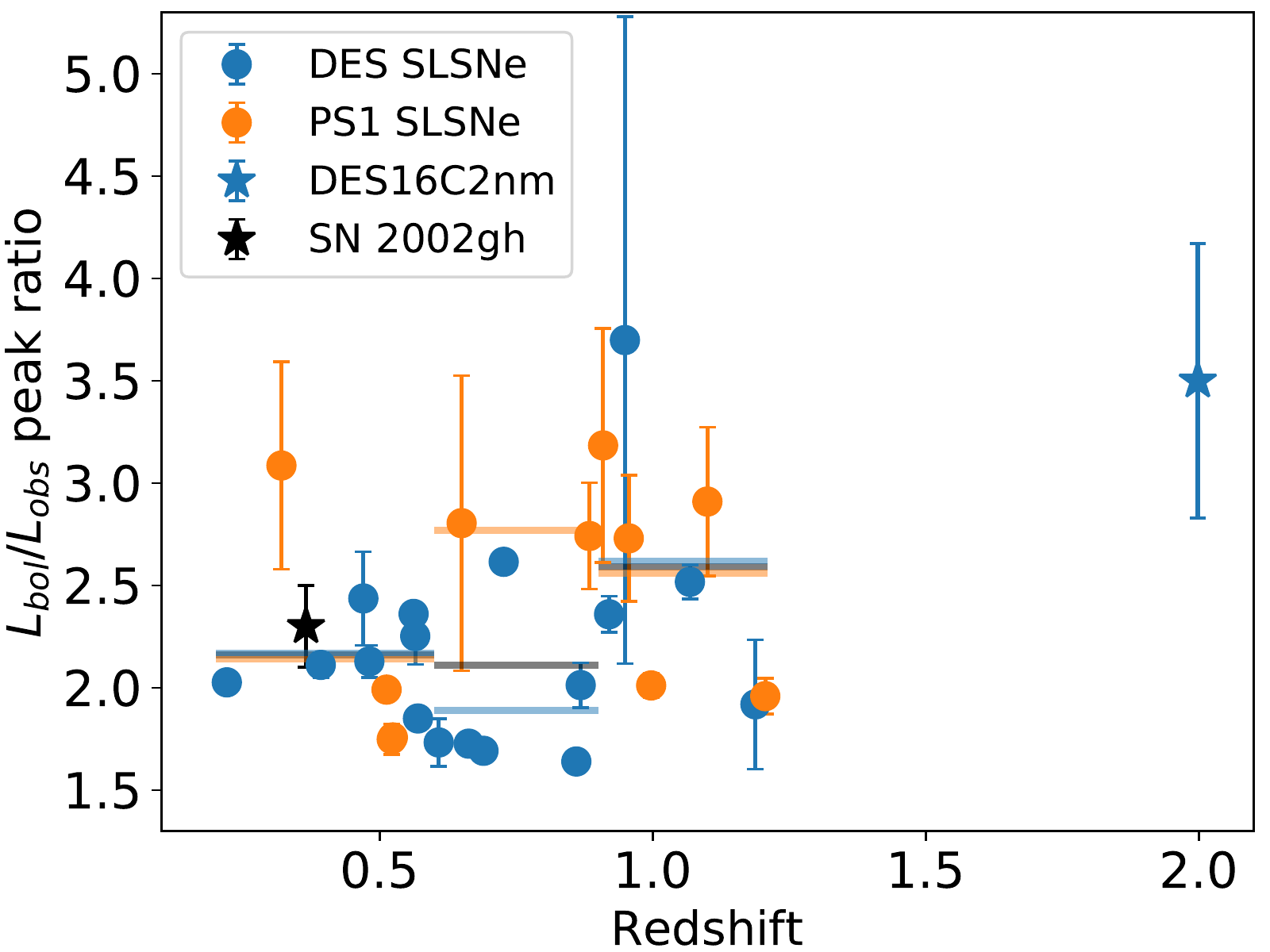}
\caption{Ratio between the maximum bolometric luminosity and the
maximum observed peak luminosity ($L_{\mathrm{bol}}/L_{\mathrm{obs}}$) as
a function of redshift. We computed the unweighted mean for $L_{\mathrm{bol}}/L_{\mathrm{obs}}$
for the redshift intervals $0.2 \textless z \textless 0.6$, $0.6 \leq z \leq 0.9$ and
$0.9 \textless z \textless 1.21$, obtaining the following values
$2.2 \pm 0.4$, $2.1 \pm 0.5$ and $2.6 \pm 0.6$, respectively.
The mean for the $L_{\mathrm{bol}}/L_{\mathrm{obs}}$ ratios are shown
as grey horizontal lines for each redshift interval. The mean for
PS1 and DES objects are shown as orange and blue horizontal lines for
the same intervals, respectively. We show the
location of SN\,2002gh and DES16C2nm in this plot with black and blue
stars, respectively. SN\,2002gh was not included in the mean computation.
}
\label{Lbol_to_Lobs_ratio_fig}
\end{figure}

In Fig. \ref{Lbol_to_Lobs_ratio_fig} the ratio between maximum $L_{\mathrm{bol}}$ and the maximum $L_{\mathrm{obs}}$ are presented. We find that the majority of
the objects have a $L_{\mathrm{bol}}/L_{\mathrm{obs}}$ value between 1.5 and 3.5, with a mean value between 2.0 and 2.5. We computed the unweighted mean
for $L_{\mathrm{bol}}/L_{\mathrm{obs}}$ for the redshift intervals $0.2 \textless z \textless 0.6$, $0.6 \leq z \leq 0.9$ and $0.9 \textless z \textless 1.21$, obtaining
the following values $2.2 \pm 0.4$, $2.1 \pm 0.5$ and $2.6 \pm 0.6$, respectively. The weighted mean gives very similar values. 
The mean values for the PS1 sample are $2.1 \pm 0.6$, $2.77 \pm 0.03$ and $2.6 \pm 0.5$
for the same redshift intervals, respectively and for the DES objects are $2.2 \pm 0.2$, $1.9 \pm 0.3$, and $2.6 \pm 0.6$ for the same intervals, respectively.
We notice that the most discrepant points are the ones with larger uncertainties, and that $L_{\mathrm{bol}}/L_{\mathrm{obs}}$ for SN\,2002gh is  $2.3 \pm 0.1$
at $z=0.365$ (we do not include SN\,2002gh in the computation).

To extend this relation beyond $z=1.2$ we used the well-observed DES16C2nm, which is a SLSN-I at $z=1.998$ \citep{smith18}. In their
study, \citet{smith18} presents optical and NIR photometry and spectroscopy, showing that DES16C2nm is spectroscopically similar to
iPTF13ajg. Unfortunately, the DES16C2nm NIR photometry covering the rest-frame optical continuum ($\lambda_{\mathrm{rest-frame}} \textgreater 3100$ \AA)
begins after maximum, therefore in order to estimate the maximum of the bolometric emission of DES16C2nm we constructed spectrophotometric templates.
To construct these templates we used the spectroscopic observations of DES16C2nm and iPTF13ajg, which were binned, smoothed and joined, and then
we scaled them using the $riz$ photometry for DES16C2nm. The DES16C2nm spectra were {\it mangled}, but did not {\it mangled} the iPTF13ajg spectra. The iPTF13ajg
spectra were firstly scaled to join to the rest-frame UV of the {\it mangled} spectra of DES16C2nm, and then the combined spectral template was re-scaled
using the $iz$ photometry. The early spectra of iPTF13ajg were scaled directly as no spectroscopic observations of DES16C2nm exists at an early phase. Then, we fitted a modified
blackbody model to the spectrophotometric templates estimating the epoch maximum bolometric luminosity on $MJD = 57684.0$, and the maximum
luminosity $L_{\mathrm{bol}} = 2.0 \pm 0.2 \times 10^{44}$ erg\,s$^{-1}$. Significant uncertanties are associated with these values due to the time gaps
between the spectrophotometric templates and the potential systematic differences between DES16C2nm and iPTF13ajg, in additon to any
other calibration issues. Therefore conservatively we quote $2.0 \pm 0.4 \times 10^{44}$ erg\,s$^{-1}$ as the estimated maximum bolometric
luminosity for DES16C2nm in Table \ref{tab:bol_DES}. Accordingly, we have extended the $L_{\mathrm{bol}}/L_{\mathrm{obs}}$ ratio to $z \simeq 2$, with 
$L_{\mathrm{bol}}/L_{\mathrm{obs}} = 3.5 \pm 0.7$ for DES16C2nm (see Fig. \ref{Lbol_to_Lobs_ratio_fig}).

Having in mind that the rest-frame wavelength and the width of the wavelength range observed in the optical bands changes dramatically over the redshift range of our sample,
we assume that this heuristic relation holds beyond $z \simeq 1.2$ and can be used to estimate the bolometric luminosity of PS1 and DES SLSNe with $z \textgreater 1.2$.
For $z \textgreater 1.2$, we assume a $L_{\mathrm{bol}}/L_{\mathrm{obs}}$ ratio of 3.0, and we set the uncertainty of the $L_{\mathrm{bol}}/L_{\mathrm{obs}}$
ratio to 0.6. Then, we compute the $L_{\mathrm{bol}}$ uncertaninty as $\sqrt{(0.6 \times L_{\mathrm{obs}})^{2}+ (3.0 \times \sigma_{L_{\mathrm{obs}}})^{2}}$.

\begin{table*}
\caption{Peak bolometric and peak observed luminosities for PS1 SLSNe.}
\label{tab:bol_PS1}
\renewcommand{\arraystretch}{1.4}
\begin{tabular}{@{}lcccccccc}
\hline
SN        & MJD peak $L_{bol}$ & Peak $L_{bol}$            & $T_{bb}$ at peak     & $R_{bb}$ at peak      & MJD peak $L_{obs}$ & Peak $L_{obs}$            & Redshift & $E(B-V)_{\mathrm{gal}}$ \\
          & (days)             & ($\times 10^{44}$ erg\,s) & ($\times 10^{3}$ K)  & ($\times 10^{15}$ cm) &                    & ($\times 10^{43}$ erg\,s) &          & \\
\hline
PS1-10ahf & 55484.4            & 0.59(0.07)                & $8.7^{+0.9}_{-0.7}$  & $3.9^{+1.0}_{-0.8}$   & 55481.35           & 2.0(0.1)                & 1.10     & 0.0293        \\
PS1-10awh & 55497.3            & 2.8(0.5)                  & $34.8^{+10.2}_{-5.7}$& $0.8^{+0.2}_{-0.2}$   & 55504.8            & 8.1(0.2)                & 0.909 & 0.0658 \\
PS1-10bzj & 55544.9            & 1.4(0.3)                  & $28.9^{+12.2}_{-5.9}$& $0.8^{+0.2}_{-0.2}$   & 55559.9            & 4.9(0.1)                & 0.650 & 0.0068 \\
PS1-10ky  & 55398.4            & 2.4(0.3)                  & $22.6^{+2.9}_{-2.2}$ & $1.5^{+0.2}_{-0.2}$   & 55408.1            & 8.9(0.2)                                 & 0.956 & 0.0299 \\
PS1-10pm  & 55334.1            & 1.17(0.04)                & $11.2^{+0.7}_{-0.6}$ & $3.4^{+0.4}_{-0.4}$   & 55323.6            & 5.9(0.2)                & 1.206  & 0.0153 \\
PS1-11afv & --                 & 2.0(0.4)$^{\ddagger}$     & --                   & --                    & 55731.4            & 6.6(0.4)                & 1.407  & 0.0138 \\
PS1-11aib & 55848.1            & 1.52(0.04)                & $15.4^{+0.6}_{-0.5}$ & $2.2^{+0.1}_{-0.1}$   & 55848.1            & 7.6(0.1)                & 0.997  & 0.0428\\
PS1-11ap  & 55612.6            & 1.42(0.03)                & $11.7^{+0.3}_{-0.3}$ & $3.4^{+0.1}_{-0.1}$   & 55612.6            & 8.0(0.1)                & 0.524  & 0.0059\\
PS1-11bam & --                 & 3.2(0.6)$^{\ddagger}$     & --                   & --                    & 55889.1            & 10.7(0.4)               & 1.565  & 0.0234\\
%PS1-11bdn & 55918.8            & 1.75(0.03)                & $17.6^{+1.4}_{-1.1}$ & $2.0^{+0.2}_{-0.2}$  & 55918.8            & 1.12(0.01)                & --           & --         \\ remove
PS1-11tt  & --                 & 1.4(0.3)$^{\ddagger}$     & --                   & --                    & 55710.6            & 4.5(0.2)                & 1.283  & 0.0080\\
PS1-12bmy & --                 & 1.5(0.3)$^{\ddagger}$     & --                   & --                    & 56206.0            & 4.9(0.2)                & 1.572  & 0.0091\\
PS1-12bqf & 56240.1            & 0.44(0.01)                & $11.6^{+0.6}_{-0.6}$ & $1.9^{+0.2}_{-0.1}$   & 56244.6            & 2.5(0.1)                & 0.522  & 0.0242\\
PS1-12cil & 56291.1            & 0.9(0.1)                  & $19.2^{+3.4}_{-1.4}$ & $1.2^{+0.1}_{-0.2}$   & 56303.1            & 2.9(0.1)                & 0.32    & 0.0234\\
PS1-13gt  & 56326.6            & 0.62(0.05)                & $7.1^{+0.4}_{-0.3}$  & $5.9^{+0.9}_{-0.8}$   & 56331.1            & 2.3(0.1)                & 0.884  & 0.0152\\
PS1-13or  & --                 & 3.8(0.8)$^{\ddagger}$     & --                   & --                    & 56393.1            & 12.6(0.2)               & 1.52   & 0.0308\\
PS1-14bj  & 56802.4            & 0.389(0.004)              & $7.4^{+0.1}_{-0.1}$  & $4.2^{+0.1}_{-0.1}$   & 56802.4            & 1.9(0.1)                & 0.5125 & 0.0189\\
\hline
\end{tabular}
\begin{tablenotes}
\item Numbers in parentheses correspond to 1\,$\sigma$ statistical uncertainties. $^{\ddagger}$Maximum bolometric luminosities estimated scaling the observed maximum luminosities.
\end{tablenotes}
%\end{center}
\end{table*}

\begin{table*}
\caption{Peak bolometric and peak observed luminosities for DES SLSNe.}
\label{tab:bol_DES}
\renewcommand{\arraystretch}{1.4}
\begin{tabular}{@{}lcccccccc}
\hline
SN         & MJD peak $L_{bol}$ & Peak $L_{bol}$            & $T_{bb}$ at peak     & $R_{bb}$ at peak       & MJD peak $L_{obs}$ & Peak $L_{obs}$            & Redshift & $E(B-V)_{\mathrm{gal}}$ \\
           & (days)             & ($\times 10^{44}$ erg\,s) & ($\times 10^{3}$ K)  & ($\times 10^{15}$ cm)  &                    & ($\times 10^{43}$ erg\,s) &          &         \\
\hline
DES13S2cmm & 56567.3            & 0.242(0.004)              & $9.5^{+0.2}_{-0.2}$  & $2.1^{+0.1}_{-0.1}$    & 56565.8            & 1.40(0.02)                & 0.663    & 0.0284  \\ 
DES14C1fi  & --                 & 2.2(0.4)$^{\ddagger}$     & --                   & --                     & 56930.9            & 7.25(0.07)                & 1.302    & 0.0091  \\
DES14C1rhg & 57008.8            & 0.196(0.001)              & $15.3^{+0.6}_{-0.5}$ & $0.78^{+0.04}_{-0.04}$ & 57013.3            & 0.92(0.01)                & 0.481    & 0.0096  \\
DES14E2slp & 57041.1            & 0.48(0.01)                & $13.1^{+0.3}_{-0.3}$ & $1.61^{+0.06}_{-0.06}$ & 57042.6            & 2.58(0.04)                & 0.57     & 0.0060  \\
DES14S2qri & --                 & 1.8(0.4)$^{\ddagger}$     & --                   & --                     & 57022.7            & 5.9(0.2)                  & 1.50     & 0.0276  \\
DES14X2byo & 56962.3            & 1.54(0.08)                & $16.6^{+0.9}_{-0.8}$ & $1.9^{+0.1}_{-0.1}$    & 56959.3            & 7.63(0.07)                & 0.868    & 0.0257  \\
DES14X3taz & 57082.6            & 1.28(0.06)                & $11.5^{+0.9}_{-0.7}$ & $3.4^{+0.5}_{-0.4}$    & 57082.6            & 7.4(0.4)                  & 0.608    & 0.0220  \\
DES15C3hav & 57337.9            & 0.209(0.006)              & $13.4^{+0.4}_{-0.3}$ & $1.03^{+0.03}_{-0.03}$ & 57340.9            & 0.99(0.01)                & 0.392    & 0.0077  \\
DES15E2mlf & --                 & 3.1(0.6)$^{\ddagger}$     & --                   & --                     & 57358.0            & 10.2(0.4)                 & 1.861    & 0.0086  \\
DES15S1nog & 57371.7            & 0.42(0.02)                & $17.9^{+1.0}_{-0.8}$ & $0.88^{+0.05}_{-0.05}$ & 57379.2            & 1.87(0.03)                & 0.565    & 0.0541  \\
DES15S2nr  & 57316.8            & 0.441(0.005)              & $10.5^{+0.1}_{-0.1}$ & $2.32^{+0.04}_{-0.04}$ & 57321.3            & 2.18(0.01)                & 0.220    & 0.0293  \\
DES15X1noe & 57450.7            & 1.5(0.2)                  & $12.6^{+3.3}_{-2.0}$ & $3.1^{+1.1}_{-0.9}$    & 57426.7            & 7.8(0.3)                  & 1.188    & 0.0177  \\
DES15X3hm  & 57239.9            & 1.55(0.02)                & $13.9^{+0.4}_{-0.4}$ & $2.6^{+0.1}_{-0.1}$    & 57235.4            & 9.46(0.09)                & 0.86     & 0.0237  \\
DES16C2aix & 57706.9            & 0.47(0.01)                & $8.4^{+0.3}_{-0.3}$  & $3.7^{+0.3}_{-0.3}$    & 57706.9            & 1.88(0.04)                & 1.068    & 0.0114  \\
DES16C2nm  & 57684.0            & 2.0(0.40)$^{\dagger}$     & $13.1^{+0.8}_{-0.8}$ & $3.5^{+0.3}_{-0.3}$    & 57632.4            & 5.96(0.02)                & 1.998    & 0.0123  \\
DES16C3cv  & 57769.9            & 0.264(0.003)              & $6.2^{+0.1}_{-0.1}$  & $5.0^{+0.1}_{-0.1}$    & 57772.9            & 1.01(0.01)                & 0.727    & 0.0103  \\
DES16C3dmp & 57723.2            & 0.65(0.01)                & $18.8^{+0.4}_{-0.4}$ & $1.00^{+0.02}_{-0.02}$ & 57726.2            & 2.76(0.02)                & 0.562    & 0.0068  \\
DES16C3ggu & 57791.6            & 1.7(0.7)                  & $29.9^{+12.7}_{-6.4}$& $0.8^{+0.2}_{-0.2}$    & 57800.6            & 4.71(0.06)                & 0.949    & 0.0074  \\
DES17C3gyp & 58152.1            & 1.78(0.02)                & $14.9^{+1.2}_{-0.9}$ & $2.5^{+0.2}_{-0.2}$    & 58144.6            & 7.3(0.2)                  & 0.47     & 0.0073  \\
DES17X1amf & 58063.9            & 1.8(0.2)                  & $18.0^{+0.7}_{-0.6}$ & $1.82^{+0.09}_{-0.09}$ & 58066.9            & 7.92(0.07)                & 0.92     & 0.0207  \\ 
DES17X1blv & 58068.2            & 0.273(0.006)              & $10.6^{+0.4}_{-0.3}$ & $1.8^{+0.2}_{-0.2}$    & 58068.2            & 1.61(0.03)                & 0.69     & 0.0213  \\
\hline
\end{tabular}
\begin{tablenotes}
\item Numbers in parentheses correspond to 1\,$\sigma$ statistical uncertainties. $^{\dagger}$Maximum bolometric luminosity estimated using spectrophotometric templates. 
$^{\ddagger}$Maximum bolometric luminosities estimated scaling the observed maximum luminosities.
\end{tablenotes}
%\end{center}
\end{table*}

% Don't change these lines
\bsp  % typesetting comment
\label{lastpage}
\end{document}